\documentclass[prd,aps,nofootinbib,superscriptaddress,floatfix,preprintnumbers,twocolumn]{revtex4-1}
\usepackage[utf8]{inputenc}
\usepackage{amsfonts}
\usepackage{bbm}
\usepackage{hyperref,amssymb,amsmath,graphicx,xcolor,slashed}
\usepackage[acronym]{glossaries}
\usepackage[normalem]{ulem}

\newacronym[longplural=parton distribution functions]{pdf}{PDF}{parton distribution function}
\newacronym{qcd}{QCD}{quantum chromodynamics}

\newcommand{\nn}{\nonumber\\}
\newcommand{\msbar}{\overline{\rm MS}}

\makeglossaries
\DeclareUnicodeCharacter{00A0}{ }

\begin{document}
\title{Direct calculation of parton distributions in momentum space from lattice QCD}

\preprint{UMD-PP-026-05, FERMILAB-PUB-26-0392-T, MIT-CTP/6043}

\author{Anthony V. Grebe}
\affiliation{Physics Department, University of Maryland, College Park, MD, 20740, USA}
\affiliation{Fermi National Accelerator Laboratory, Batavia, IL 60510, USA}

\author{Daniel C. Hackett}
\affiliation{Fermi National Accelerator Laboratory, Batavia, IL 60510, USA}

\author{Michael L. Wagman}
\affiliation{Fermi National Accelerator Laboratory, Batavia, IL 60510, USA}

\author{Rui Zhang}
\email{rzhang93@mit.edu}
\affiliation{Center for Theoretical Physics - a Leinweber Institute, Massachusetts Institute of Technology, Cambridge, MA 02139, USA}
\affiliation{Physics Division, Argonne National Laboratory, Lemont, IL 60439, USA}

\author{Yong Zhao}
\affiliation{Physics Division, Argonne National Laboratory, Lemont, IL 60439, USA}

\begin{abstract}
Coulomb-gauge quasi-parton distributions can be computed directly in momentum space on a finite lattice, enabled by the commutativity of their renormalization and Fourier transform. This approach removes the formal inverse problem in coordinate-space methods.
Our momentum-space pion quasi-distributions agree with coordinate-space results Fourier transformed with asymptotic extrapolation, indicating that the formal inverse problem in the latter is not a concern at this volume.
We further extend the framework to higher dimensions and obtain the first 3D image of the pion directly from lattice QCD.
\end{abstract}

\maketitle

Parton distribution functions (PDFs) describe the confined motion of quarks and gluons inside hadrons and serve as key nonperturbative inputs for precision Standard Model predictions in high-energy scattering experiments. They lie at the core of the scientific programs of facilities such as Jefferson Lab~\cite{Dudek:2012vr}, the Electron-Ion Collider~\cite{Accardi:2012qut,AbdulKhalek:2021gbh}, and the Large Hadron Collider~\cite{Apollinari:2015wtw}.
Apart from global fits in phenomenology, lattice QCD provides a first-principles approach to determine PDFs~\cite{Constantinou:2020hdm}. Since PDFs are defined in terms of lightlike correlations with real-time dependence, they cannot be computed directly on a Euclidean lattice with imaginary time.
To overcome this hurdle, various methods have been developed~\cite{Kronfeld:1984zv,Martinelli:1987si,Davoudi:2012ya,Shindler:2023xpd,Braun:2007wv,Radyushkin:2017cyf,Ma:2017pxb,Liu:1993cv,Detmold:2005gg,Detmold:2021uru,Chambers:2017dov,Ji:2013dva,Ji:2014gla,Ji:2020ect}, which can be broadly grouped into three categories: Mellin moments~\cite{Kronfeld:1984zv,Martinelli:1987si,Davoudi:2012ya,Shindler:2023xpd}, short-distance factorization (SDF)~\cite{Braun:2007wv,Radyushkin:2017cyf,Ma:2017pxb} or operator product expansion (OPE)~\cite{Detmold:2005gg,Detmold:2021uru,Chambers:2017dov}, and Large Momentum Effective Theory (LaMET)~\cite{Ji:2013dva,Ji:2014gla,Ji:2020ect}. Among them, LaMET determines the Bjorken-$x$ dependence of PDFs through a systematic power expansion and effective-theory matching of Euclidean quasi-PDFs in a highly boosted hadron state~\cite{Ji:2020byp,Ji:2024oka}. This framework can be further extended to 3D structures~\cite{Ji:2020ect}, including generalized parton distributions (GPDs) and, in a distinct manner, transverse-momentum-dependent distributions (TMDs).

In the past decade, substantial progress has been made in lattice calculations of parton physics~\cite{Lin:2025hka}, particularly within the LaMET framework~\cite{Zhao:2025oto}. Uncertainty quantification has thus become an important {topic}, attracting considerable attention from both the lattice and phenomenology communities~\cite{INT24-88W,PDFLattice2024,DelDebbio:2024sfa,Dutrieux:2025axb,Chen:2025cxr,Xiong:2025obq,Chu:2025jsi,Dutrieux:2025jed,Ling:2025olz,Rothkopf:2026wdj}.
In all approaches developed so far, one must reconstruct a smooth distribution in $x$-space from a finite set of correlation-function data points, {which is formally an inverse problem (IP) if there is no extra constraint}.
While the SDF and OPE methods need to fit the PDFs from a short segment of coordinate-space correlations or the lowest Mellin moments, LaMET requires the Fourier transform (FT) over a broad range of spatial correlations to obtain the quasi-PDFs. Due to the finite lattice extent and increasing statistical noise, the FT in LaMET typically has to be truncated at distances of about $1.0$~fm.
Nevertheless, confinement in QCD constrains the asymptotic behavior of spatial correlations at large separations~\cite{Ji:2020brr,Chen:2025cxr,Gao:2021dbh,Ji:2026vir}. These constraints can be rigorously derived and used to extrapolate the correlations to infinite separation~\cite{Ji:2026vir}, thereby enabling a controlled FT with systematic uncertainty quantification.

{The question of how to quantify FT uncertainties does not arise in direct calculations of momentum-space quasi-distributions, where the formal IP is absent by construction.}
However, this is impractical for the original quasi-PDFs defined from Wilson-line operators~\cite{Ji:2013dva}, since their ultraviolet linear divergences depend on the separation~\cite{Dotsenko:1979wb,Craigie:1980qs,Dorn:1986dt,Ji:2017oey,Ishikawa:2017faj,Green:2017xeu} and must be removed through renormalization prior to the FT.
Recently, it has been proposed that Coulomb-gauge (CG) correlators~\cite{Gao:2023lny,Zhao:2023ptv}, which belong to the universality class~\cite{Hatta:2013gta,Ji:2020ect} of quasi-observables in LaMET, can also be used to compute parton physics. Thanks to the absence of Wilson lines and linear divergences~\cite{Gao:2023lny,Zhang:2024omt}, the CG approach substantially improves the precision of large-separation observables like TMDs~\cite{Zhao:2023ptv,Bollweg:2024zet,Bollweg:2025iol,Bollweg:2025ecn}.

In this Letter, we point out that, since the renormalization is independent of the separation, the discrete FT (DFT) of the CG correlator can be directly simulated on a finite lattice as a momentum-space operator.
{In this way, no formal IP enters this approach.}
In addition, this method enables the direct computation of 3D momentum-space observables, e.g., TMDs, which has not been feasible {in lattice QCD. Existing calculations are typically performed in transverse-coordinate space and require a model-dependent FT to reconstruct the corresponding momentum-space images~\cite{Bollweg:2025iol}}.

Using this method, we construct momentum-space CG quasi-PDF operators on the lattice and directly simulate them in a boosted pion state. We also compute the coordinate-space matrix elements parasitically using the same lattice setup. The valence quark quasi-PDFs show very good agreement with those obtained from the FT of the coordinate-space matrix elements with asymptotic extrapolation~\cite{Ji:2026vir}, including at fractional lattice momenta beyond discrete Fourier modes.
We then renormalize the quasi-PDFs in the $\msbar$ scheme and match them to the lightcone PDFs at next-to-leading order (NLO), which again show good agreement with those from the conventional LaMET analysis. Our results cross-validate both approaches in handling the FT in LaMET. Moreover, we calculate the valence quark quasi-TMDs in momentum space and, for the first time, obtain a 3D image of the pion directly from the lattice.\\

\paragraph*{Method.}
Traditionally, the quasi-PDF is defined from the following matrix element,
\begin{align}
    \tilde{h}(z,P_z)={1\over 2P^t}\langle P|\bar{\psi}(0){\gamma^t}W(0,z)\psi(z)|P\rangle,
\end{align}
where $|P\rangle$ is a plane-wave hadron state with momentum $P^\mu=(P^t,0,0,P^z)$, and $W(0,z)$ is a Wilson line that makes the non-local operator gauge invariant.
On a lattice with spacing $a$, the bare matrix element $\tilde{h}^B(z,P_z,a)$ is renormalized and converted to the $\overline{\rm MS}$ quasi-PDF $\tilde{q}^R(x,P_z,\mu)$ through FT,
\begin{align}
    \tilde{q}^R(x,P_z,\mu)=P_z\int_{-\infty}^\infty \frac{dz}{2\pi} e^{ixzP_z} \frac{\tilde{h}^B(z,P_z,a)}{Z(z,a,\mu)}\,,
\end{align}
where the superscripts $B$ and $R$ represent bare and renormalized quantities.
The renormalization factor $Z(z,a,\mu)$ contains a linear divergence $\exp(-\delta m(a)|z|)$ with $\delta m(a) \sim 1/a$~\cite{Dotsenko:1979wb,Craigie:1980qs,Dorn:1986dt,Ji:2017oey,Ishikawa:2017faj,Green:2017xeu}, which must be subtracted prior to the FT. Due to finite lattice size and increasing statistical noise, the range of $z$ is usually limited in the FT, which requires an extrapolation of the long tail to infinity.
The recent works in Refs.~\cite{Dutrieux:2025axb,Chen:2025cxr,Xiong:2025obq,Chu:2025jsi,Dutrieux:2025jed,Ling:2025olz} centered on how to do the FT and quantify the associated uncertainties.
Fortunately, due to color confinement, $\tilde{h}(z,P_z)$ decays exponentially at large $|z|$, whose asymptotic form has been rigorously derived~\cite{Ji:2026vir}. Therefore, one can perform a physics-motivated extrapolation to $|z|\to\infty$ to complete the FT with controlled uncertainties.

Actually, the DFT of a periodic observable on a finite lattice is unique and can be explicitly computed without any formal IP.
For Wilson-line operators, the linear divergence breaks periodicity and obstructs direct momentum-space calculations.
However, there exist alternative quasi-observables within the LaMET universality class~\cite{Hatta:2013gta,Ji:2020ect} that are free from this issue, such as the CG correlators~\cite{Gao:2023lny,Zhao:2023ptv},
\begin{align}
    O_C(z)&= \bar{\psi}(0)\gamma^t\psi(z)\Big|_{\vec{\nabla}\cdot \vec{A}=0} \,.
\end{align}
where $\vec{\nabla}\cdot \vec{A}=0$ is the CG condition.

{There has been substantial progress toward a complete proof of the renormalizability of CG QCD in the literature~\cite{Zwanziger:1998ez,Baulieu:1998kx,Niegawa:2006ey,Niegawa:2006hg}. Meanwhile, lattice calculations have shown that equal-time quark correlators are multiplicatively renormalizable and free from linear divergences~\cite{Burgio:2012ph,Gao:2023lny,Zhang:2024omt},}
\begin{align}\label{eq:ren}
    O_C^R(z,\mu)=O_C^B(z,a)/Z^R_C(a,\mu)\,,
\end{align}
where $Z^R_C(a,\mu)$ is the $z$-independent CG quark wave function renormalization factor.
Lattice quasi-PDFs can therefore be computed directly in momentum space using a DFT of CG correlators.
At finite lattice size $L$, one can define a momentum-space operator
\begin{align}
    \tilde O_C(k_z)&\equiv\sum_{z=-L/2}^{L/2-a}  e^{izk_z} \bar{\psi}(0)\gamma^t\psi(z),
\end{align}
which remains multiplicatively renormalizable as Eq.~\eqref{eq:ren}.
Here $\tilde O_C(k_z)$ is proportional to the number density operator $\psi^\dagger(k^z)\psi(k^z)$ when sandwiched between hadron states, which directly gives the quasi- PDFs and GPDs.

Moreover, it can be generalized to 3D space as
\begin{align}\label{eq:tmd_op}
    \tilde O_C(\vec{k})&\equiv\sum_{\vec{b}\in V}  e^{i\vec{k}\cdot \vec{b}} \bar{\psi}(0)\gamma^t\psi(\vec{b})\,,
\end{align}
with $\vec{k}=(\vec{k}_T, k_z)$ and $\vec{b}=(\vec{b}_T, b_z)$, whose matrix element defines the quasi-TMD,
\begin{align}\label{eq:quasi_mtm}
    \tilde q(x, \vec{k}_T, P_z) &= {P_z\over 2P^t}\langle P|\tilde O_C(\vec{k})|P\rangle\,,
\end{align}
and quasi-Wigner distribution {in which the initial and final states carry different momenta~\cite{Belitsky:2003nz}}.
Since $\tilde O_C(\vec{k})$ has a physical---albeit frame-dependent---interpretation as the momentum-density operator $\psi^\dagger(\vec{k})\psi(\vec{k})$ between hadron states~\cite{Ji:2020ect}, it enables direct 3D imaging on the lattice without modeling the $\vec{b}_T$ dependence of coordinate-space matrix elements~\cite{Bollweg:2025ecn}.~\footnote{To match $\tilde q(x, \vec{k}_T, P_z)$ to the TMDPDF appearing in TMD factorization theorems~\cite{Collins:2011zzd}, a soft-function subtraction remains necessary~\cite{Ji:2019ewn,Zhao:2023ptv}, which can be implemented as a convolution over the first Brillouin zone in $\vec{k}_T$-space.}

The resulting quasi-distributions are periodic functions in $x=k_z/P_z$ with a period of $2\pi/(aP_z) \gg 1$, which typically decay rapidly to zero as $x$ surpasses $1$.
The hadron momentum $P_z$ must match the exact Fourier modes $n({2\pi}/{aL})$ with integer $n$. When $k_z$ matches the integer Fourier modes, the result is exact; otherwise, the FT breaks the periodicity and introduces extra finite-volume effects. Fortunately, for correlation functions that fall off fast within $|z|<L/2$, such effects are suppressed. In particular, since they exhibit an asymptotic exponential decay $e^{-\Lambda |z|}$, one can estimate the finite-volume correction to be
\begin{align}\label{eq:fv_bound_2}
    |\Delta\tilde{q}_{\rm lat}(x)|&\sim 2e^{-\Lambda L}|\sin(xLP_z/2)||\tilde{q}(x)|\,,
\end{align}
where $\tilde{q}(x)$ is the infinite volume quasi-distribution.
(See Appendix~\ref{app:fv}).
{For example}, if $\Lambda > m_\pi$ and $m_\pi L>4$, then $2e^{-\Lambda L}<0.04$, which is typically negligible compared with other uncertainties and can be systematically quantified through the infinite-volume extrapolation.

\paragraph*{Numerical results.}
We follow this strategy to calculate the pion valence PDF on a gauge ensemble produced by the MILC collaboration~\cite{MILC:2012znn}, with 2+1+1 flavors of highly improved staggered quarks (HISQ) tuned to pion mass $m_\pi\approx 310$~MeV and the one-loop Symanzik improved gauge action~\cite{Symanzik:1983dc}. The lattice volume and spacing are $L^3\times T=48^3\times144$ and $a\approx0.06$~fm. We apply one step of HYP smearing with parameters $\{\alpha_1,\alpha_2,\alpha_3\}=\{0.75,0.6,0.3\}$~\cite{Hasenfratz:2001hp} to the gauge links and use a Wilson-clover action~\cite{Sheikholeslami:1985ij} for the valence quarks with
$c_\text{SW} = 1.03493$ and bare light quark mass $m_q=-0.0398$,
tuned to produce the valence pion mass $m_\pi\approx 310$~MeV. The gauge ensembles are fixed to Coulomb gauge condition with precision $10^{-14}$~\cite{Giusti:2001xf,Hudspith:2014oja}.

To increase the signal at large momentum, we use the kinematically enhanced interpolators~\cite{Zhang:2025hyo} for pion,
\begin{align}
    \chi_\pi=\bar{d}\gamma_+\gamma_5u,
\end{align}
which also reduces the computing time by half since only $6$ out of 12 spin-color indices at the source and the sink are used during the smearing and inversion~\cite{Reitinger:2026hta}. We also apply momentum smearing~\cite{Bali:2016lva} with $k\approx1.7$~GeV and anisotropic Gaussian width $\vec{\sigma}=(0.48,0.48,0.18)$~fm for pion momentum $P_z=\{ 2.2, 2.6\}$~GeV. The anisotropic smearing significantly reduces excited-state contamination~\cite{keli2026}. The measurements are performed across 100 gauge configurations with 16 sources each.

We measure the following two-point and three-point correlation functions,
\begin{align}
    C_{\rm 2pt}(\tau,P_z)&=\sum_{\vec{y}\in V}e^{i\vec{y}\cdot \vec{P}}
   \langle \chi(y,\tau)\bar{\chi}(0,0)\rangle,\\
    C_{\rm 3pt}(t,\tau,xP_z)&=\sum_{\vec{y}\in V}e^{i\vec{y}\cdot \vec{P}}\langle \chi(\vec{y},\tau)\tilde{O}_C(xP_z,t)\bar{\chi}(0,0)\rangle,\nonumber
\end{align}
where $V$ is the spatial lattice volume and $\chi$ is the hadron interpolator.
The quasi-PDF matrix elements are extracted from their ratios {at $\tau\gg t\gg a$},
\begin{align}
    R(x,P_z)&\equiv \frac{C_{\rm 3pt}(t,\tau,xP_z)}{C_{\rm 2pt}(\tau,P_z)} \\
    &=\langle P|\tilde O_C(xP_z)|P\rangle +\mathcal{O}(e^{-\Delta E(\tau-t)},e^{-\Delta Et})\,,\nonumber
\end{align}
where $\Delta E$ is the energy gap between the ground state and excited states.

\begin{figure}
    \centering
    \includegraphics[width=0.9\linewidth]{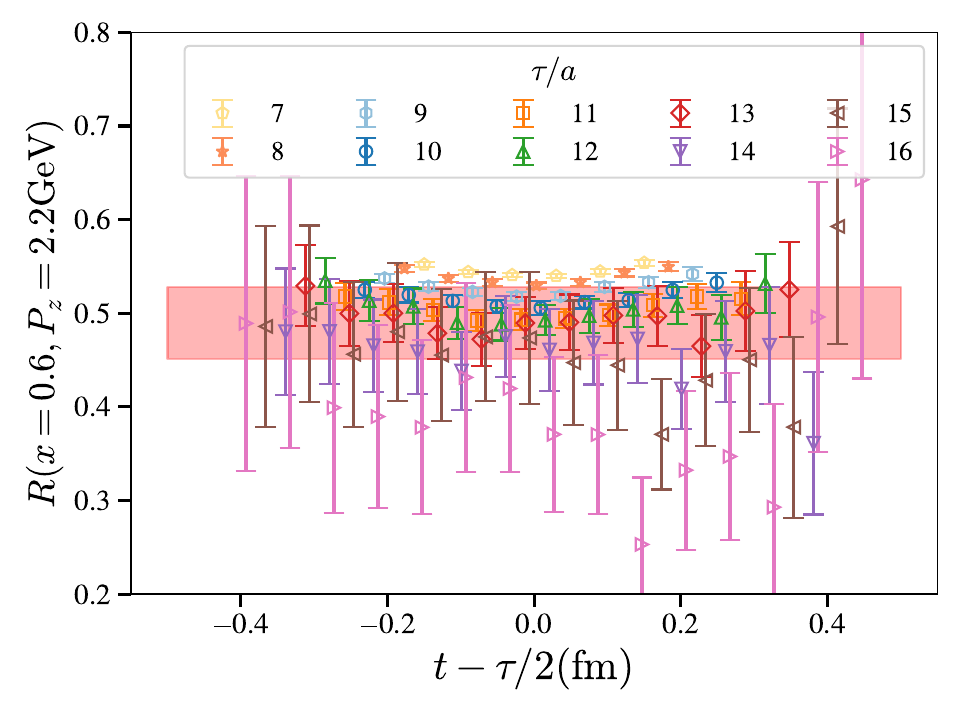}
    \caption{The $C_{\rm 3pt}/C_{\rm 2pt}$ ratios for the momentum-space operator, with the ground-state bare matrix element (in red bands) extracted with the Lanczos method~\cite{Hackett:2024xnx}.}
    \label{fig:3pt}
\end{figure}

We analyze $C_{\rm 2pt}$ with the Lanczos method to extract the ground-state energy and Ritz vectors~\cite{Wagman:2024rid}, which are then applied to $C_{\rm 3pt}$ to extract the ground-state matrix elements~\cite{Hackett:2024xnx}. For a robust analysis, we generate $C_{\rm 3pt}(t,\tau,xP_z)$ with $\tau\in[2,17]a$, such that up to 8 iterations are allowed. For traditional multi-state analysis, only a few large $\tau$ values are needed. The results for $R(x,P_z)$ at $x=0.6$ and $P_z=2.2$ GeV are shown in Fig.~\ref{fig:3pt}. We find that both the ground-state energy and $R(x,P_z)$ converge nicely to the extracted values at large Euclidean time, which indicates that the excited-state contamination has been controlled well. Besides, {the $C_{\rm 3pt}/C_{\rm 2pt}$ ratios in $z$-space also} demonstrate good convergence towards the ground-state matrix elements.

\begin{figure}
    \centering
    \includegraphics[width=0.95\linewidth]{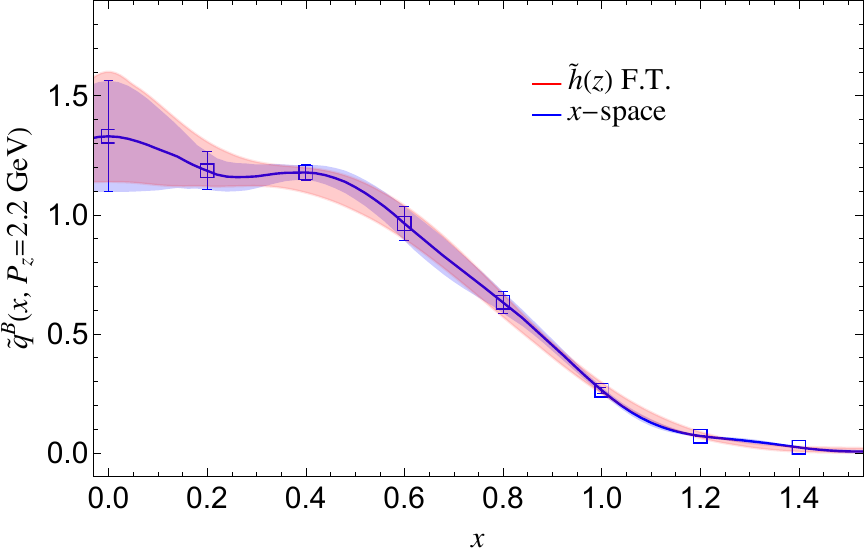}
    \caption{The quasi-PDF obtained from asymptotic analysis of coordinate-space matrix elements (labeled as ``$\tilde{h}(z)$ F.T.'') compared to the direct measurements in $x$-space (labeled as ``$x$-space''). The blue squares represent integer Fourier modes.}
    \label{fig:compare_ft}
\end{figure}

Figure~\ref{fig:compare_ft} shows the bare quasi-PDF (labeled as ``$x$-space'') calculated directly in momentum space.
The blue squares are integer Fourier modes, while the blue band corresponds to non-integer modes, which contain finite-volume effects bounded by Eq.~\eqref{eq:fv_bound_2}. Notably, the latter is smoothly connected to the integer modes, indicating that the finite-volume effects are small.

Furthermore, to compare with the asymptotic extrapolation in traditional LaMET analysis~\cite{Ji:2020brr}, we have also extracted the coordinate-space matrix elements $\tilde{h}(z,P_z,a)$. At $|z|>0.5$~fm, $\tilde{h}(z)$ has the following expansion up to the next-to-leading asymptotic order~\cite{Ji:2026vir,Gao:2026hix},
\begin{align}\label{eq:asymp}
    \tilde{h}(z\to\infty)=z^{-\alpha}\left(A_0e^{i\phi_0}+\frac{A_1 e^{i\phi_1}}{|z|}\right)e^{-|z|M} + \dots,
\end{align}
where $\alpha$, $A_{0,1}$, $\phi_{0,1}$ and $M$ are fitting parameters, and {$\alpha$ is {left as a free fit parameter} to account for CG dynamics}~\cite{Gao:2026hix}. {Then we obtain the $x$-space quasi-PDF by performing a DFT of the lattice data within $|z|\leq z_{\rm cut}= 0.54$~fm, plus a continuous integral of the fitted asymptotic form for $z\in[z_{\rm cut},\infty]$.
We also test the leading asymptotic expansion by setting $A_1=0$ and fitting from larger $z_{\rm cut}=0.78$~fm. Both fits yield $\chi^2/{\rm d.o.f}<0.2$.} The model variance is included in the error of the $x$-space results after FT, shown as the red band in Fig.~\ref{fig:compare_ft}.
{The results from both approaches show good agreement, although the coordinate-space method exhibits slightly larger uncertainties. This provides a nontrivial cross-validation of the two methods for handling the FT in LaMET }
{and verifies the formal IP is not a practical concern for these results.}

\begin{figure}
    \centering
    \includegraphics[width=0.99\linewidth]{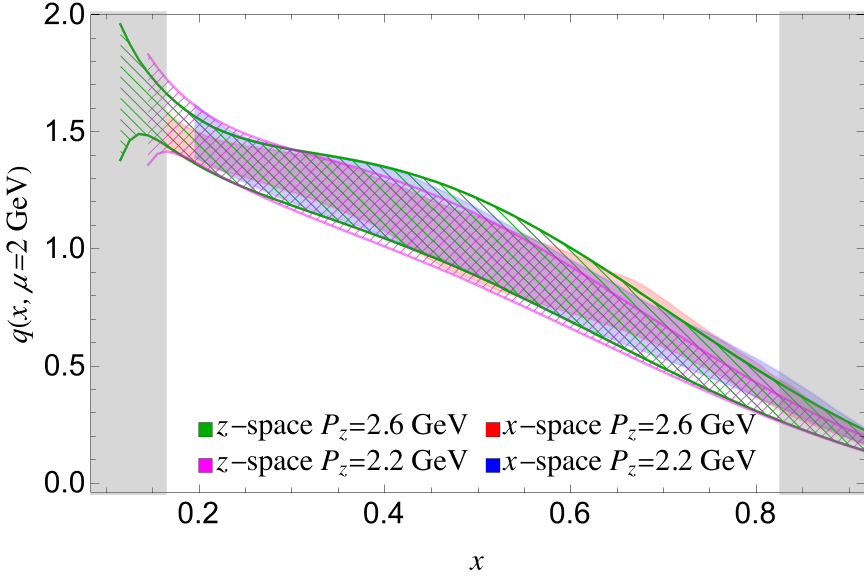}
    \caption{Comparison between the lightcone PDFs at $\msbar$ scale $\mu=2$~GeV obtained from the direct $x$-space method (labeled as ``$x$-space'') and traditional analysis of coordinate-space matrix elements (labeled as ``$z$-space''). The two methods yield consistent results at different pion momenta. The shaded regions, $x<x_0$ and $x>1-x_0$ with $x_0\approx0.18$, indicate where perturbation theory becomes unreliable.}
    \label{fig:compare_x}
\end{figure}

Next, we renormalize the quasi-PDF $\tilde q^B(x,P_z,a)$ in the $\msbar$ scheme following the procedure in Ref.~\cite{Gao:2026hix}.
The renormalization factor is precisely calculated on the same lattice as $Z_C^{\msbar}=0.9215(4)$.
See Appendix~\ref{app:renorm} for more details.
Then we match the quasi-PDF to the lightcone through the LaMET expansion,
\begin{align}
    q(x,\mu)=\int& {dy\over |y|} C^{\msbar}\bigg(\frac{x}{y},xP_z,\mu\bigg)\tilde{q}^{\msbar}(y,P_z,\mu) \nn
    &+\mathcal{O}\bigg(\frac{\Lambda^2_{\rm QCD}}{(xP_z)^2},\frac{\Lambda^2_{\rm QCD}}{((1-x)P_z)^2}\bigg)
\end{align}
where $C^{\msbar}\left({x}/{y},xP_z,\mu\right)$ is the $\msbar$ matching kernel known at NLO~\cite{Gao:2023lny}, plus an $\mathcal{O}(a)$ correction term as discussed in Appendix~\ref{app:oa_corr}.
To include higher-order logarithmic corrections, we perform the renormalization group (RG) resummation by matching at $\mu=2\kappa xP_z$ and evolving back to $2$~GeV for each $x$~\cite{Su:2022fiu}, and include the scale variation $\kappa \in[1/\sqrt{2},\sqrt{2}]$ in our systematic error. We shade the endpoint regions $x\leq x_0\approx0.18$ and $x\geq 1-x_0$ where $\alpha_s(2\kappa x_0P_z)>0.5$ and perturbation theory becomes unreliable~\cite{Ji:2024hit}. The results are then compared with those obtained from the traditional analysis with hybrid-scheme renormalization and asymptotic extrapolation~\cite{Gao:2023lny}. Although the two approaches use different renormalization schemes, the final results are consistent within error, as shown in Fig.~\ref{fig:compare_x}. The results at $P_z=2.2$~GeV and $P_z=2.6$~GeV are also consistent within both approaches, indicating small power corrections.
Notably, our results are consistent with the previous calculation using CG correlators in Ref.~\cite{Gao:2023lny}, where consistency with traditional gauge-invariant correlators was also demonstrated.

\begin{figure}
    \centering
    \includegraphics[width=1.0\linewidth]{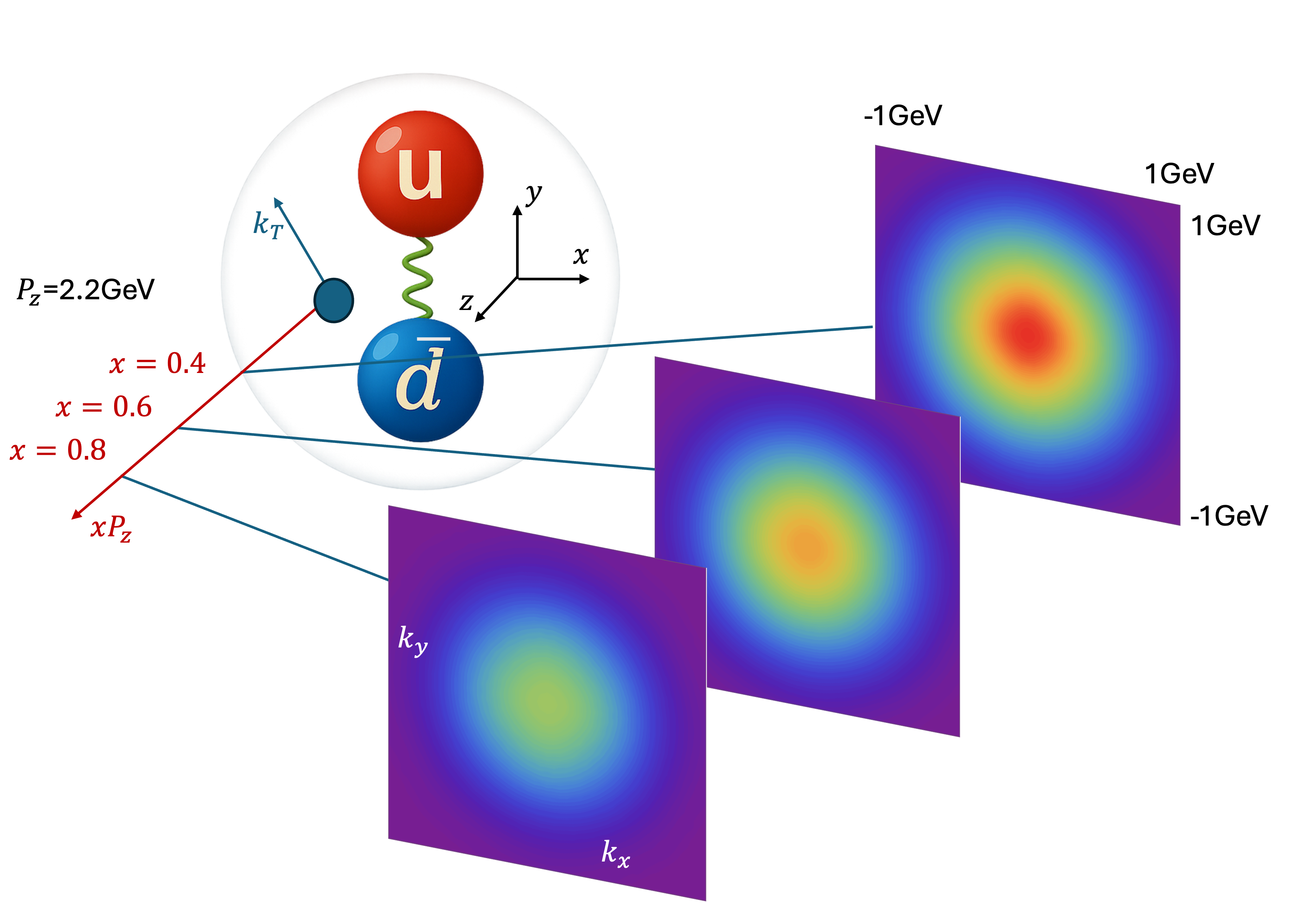}
    \caption{The 3D momentum density distribution of a valence quark inside a moving pion at $P_z=2.2$~GeV.}
    \label{fig:pion_3d}
\end{figure}

Finally, using the generalized formalism in Eq.~\eqref{eq:tmd_op}, we measure this 3D momentum density of the valence quark in a moving pion with $P_z=2.2$~GeV, as shown in Fig.~\ref{fig:pion_3d}.
{This represents the first direct lattice calculation of a 3D image of a hadron.}
{The overall $\vec{k}_T$-distribution is more suppressed at larger $x$, which}
is similar to that observed in the physical TMDPDFs from the JAM23 global fit~\cite{Barry:2023qqh} and the lattice calculation of Ref.~\cite{Bollweg:2025iol}, but differs from AV19~\cite{Vladimirov:2019bfa} and MAP22~\cite{Cerutti:2022lmb}, where {the shape of} the distributions become sharper and broader, respectively, when $x$ increases. Note that the $\vec{k}_T$-distribution we calculate is frame dependent, whose $P_z$-dependence is described by the Collins-Soper evolution equation~\cite{Collins:2011zzd,Zhao:2023ptv}.
By calculating the distribution at a different pion momentum, we find that it becomes broader in $k_T$ as $P_z$ increases, {agrees with Collins-Soper evolution.} More details are provided in Appendix~\ref{app:cs_evo}.

\paragraph*{Conclusion.}
In this work, we have proposed to directly simulate CG quasi-parton distributions in momentum space on the lattice{, which removes the formal IP in the FT of coordinate-space methods}.
Thanks to the asymptotic exponential decay of CG correlators, the momentum-space distributions can be evaluated at fractional lattice momenta beyond integer Fourier modes, with finite-volume effects that can be systematically controlled.
Through a numerical calculation of the pion valence PDF, we demonstrate that this method yields $x$-space results consistent with the traditional LaMET analysis of coordinate-space matrix elements, thereby providing a cross-validation of both approaches for handling the FT. Moreover, by calculating momentum-space quasi-TMDs, we obtain for the first time a 3D image of the pion directly from lattice QCD, whose evolution in $P_z$ agrees with the Collins-Soper equation.

{Our momentum-space approach provides a new and simple framework within LaMET. Moving forward, it can be combined with coordinate-space methods for cross-validation of the results at essentially no additional computational cost, thereby further advancing precision lattice calculations of parton physics. Furthermore, it enables model-independent 3D imaging of the nucleon, which will greatly complement the current and future experiments.}

\section*{Acknowledgment}
We thank Xiang Gao, Joseph Karpie, Andreas Kronfeld, and Yushan Su
for helpful discussions on discrete Fourier transform and inverse problems.
We also thank Xiang Gao, Yang Fu, Jinchen He, Xiangdong Ji, Phiala Shanahan, and Yushan Su
for other helpful suggestions and discussions.
We thank the MILC collaboration for sharing the gauge ensembles. The HYP smearing and Coulomb gauge fixing are implemented with the Gauge Link Utility (GLU) library~\cite{Hudspith:2014oja}. The measurement of the lattice correlators are performed with the PyQUDA software~\cite{Jiang:2024lto} based on the QUDA library~\cite{Clark:2009wm}. Computations for this work were carried out in part on facilities of the USQCD Collaboration, which is funded by the Office of Science of the U.S. Department of Energy. This research also used resources of the National Energy Research
Scientific Computing Center, a DOE Office of Science User Facility
supported by the Office of Science of the U.S. Department of Energy
under Contract No. DE-AC02-05CH11231 using NERSC award
NP-ERCAP0037091, and DELTA at UIUC through allocation PHY250232 from the Advanced Cyberinfrastructure Coordination Ecosystem: Services \& Support (ACCESS) program, which is supported by U.S. National Science Foundation grants \#2138259, \#2138286, \#2138307, \#2137603, and \#2138296. YZ is supported by the U.S. Department of Energy, Office of Science, Office of Nuclear Physics through Contract No.~DE-AC02-06CH11357, and the Early Career Award through Contract No.~DE-SCL0000017. RZ is supported
by the U.S. Department of Energy, Office of Science, Office of Nuclear Physics under grant Contract No.~DE-SC0011090 and DOE Quark-Gluon Tomography (QGT) Topical Collaboration under award No.~DE-SC0023646.

\appendix

\section{Extra finite volume effects for non-integer Fourier modes}
\label{app:fv}
Because the lattice propagators and gauge fields obey periodic boundary conditions, the matrix elements $\tilde h(z)$ measured on a finite lattice can be interpreted as a sum of non-periodic correlation $H(z+nL)$ from all periodic images, i.e., from all integers $n$, in the infinite-volume space filled by periodically repeated images,
\begin{align}
    \tilde h(z)= \sum_n H(z+nL).
\end{align}
Thus, the FT is
\begin{align}\label{eq:periodic_general}
    \tilde{q}_{\rm lat}(x)&\equiv \sum_{z=-L/2}^{L/2-a}\frac{aP_z}{2\pi} e^{ixzP_z} \tilde h(z)\nn
    &=\sum_n\sum_{z=-L/2}^{L/2-a}\frac{aP_z}{2\pi} e^{ixzP_z} H(z+nL).
\end{align}
When the parton momentum is $xP=k\times \frac{2\pi}{L}$ with an integer $k$, the phases from different periods are the same, thus it is simplified to
\begin{align}\label{eq:periodic_int}
    \tilde{q}_{\rm lat}(x)
    &=\tilde{q}(x)\equiv\sum_{z=-\infty}^{\infty}\frac{aP_z}{2\pi} e^{i2\pi kz/L}H(z),
\end{align}
which matches the infinite-volume results $\tilde{q}(x)$ exactly. But when we calculate at non-integer modes, there will be a deviation in the phase, resulting in a difference between $\tilde{q}(x)$ and $\tilde{q}_{\rm lat}(x)$,
\begin{align}
    \Delta\tilde{q}_{\rm lat}(x)
    &=\sum_n\sum_{z=-L/2}^{L/2-a}\frac{aP_z}{2\pi} e^{ixzP_z}[ 1-e^{ixnLP_z}]H(z+nL).
\end{align}

To estimate the deviation, we utilize the fact that pion correlations of valence quarks decay exponentially in the non-trivial QCD vacuum {and are symmetric in $z$},
\begin{align}
    H(|z|+L)\sim e^{-\Lambda L} H(|z|), \qquad H(z)=H(-z),
\end{align}
then the deviation can be simplified as
\begin{align}
    &\Delta\tilde{q}_{\rm lat}(x)
    =\sum_{z=-L/2}^{L/2-a}\frac{aP_z}{2\pi} e^{ixzP_z}\sum_{n\neq 0}[ 1-e^{ixnLP_z}]H(z+nL)\nn
    &\approx\sum_{z=-L/2}^{L/2-a}\frac{aP_z}{2\pi} e^{ixzP_z}\sum_{n> 0}[ 1-e^{ixnLP_z}]\nonumber\\
    &\qquad \times [H(z)+H(L-|z|)]e^{-n\Lambda L} \nn
    &=\sum_{z=-L/2}^{L/2-a}\frac{aP_z}{2\pi} e^{ixzP_z} H(z)(1+\mathcal{O}(e^{-\Lambda L/2}))\nn
    &\qquad\times\frac{e^{-\Lambda L}(1-e^{ixLP_z})}{(1-e^{-\Lambda L})(1-e^{ixLP_z-\Lambda L})}\nn
    &\approx \frac{\tilde{q}(x)e^{-\Lambda L}(1-e^{ixLP_z})}{1-e^{-\Lambda L}}(1+\mathcal{O}(e^{-\Lambda L/2}))\,,
\end{align}
where the $H(L-|z|)$ terms becomes $\mathcal{O}(e^{-\Lambda L/2})$, and we have approximated $\tilde q(x)$ as
\begin{align}
    \tilde{q}(x)&\approx \sum_{z=-L/2}^{L/2-a}\frac{aP_z}{2\pi} e^{ixzP_z} H(z)\sum_n e^{-n\Lambda L-ixnLP_z}\nn
    &=\sum_{z=-L/2}^{L/2-a}\frac{aP_z}{2\pi} e^{ixzP_z} H(z)\frac{1}{(1-e^{ixLP_z-\Lambda L})},
\end{align}
in the last step.
In the limit $\Lambda L\gg1 $, the relative deviation $|\Delta\tilde{q}_{\rm lat}(x)|/|\tilde{q}(x)|$ scales as $e^{-\Lambda L}$, which has a similar form to the conventional finite volume effects.
If we assume $\Lambda \gtrsim m_\pi$, {as indicated by our fitted $M\approx0.6$~GeV in Eq.~\eqref{eq:asymp},}
for a typical lattice with $m_\pi L\sim4$,
\begin{align}
\frac{|\Delta\tilde{q}_{\rm lat}(x)|}{|\tilde{q}(x)|}\approx 2|\sin (xLP_z/2)|e^{-\Lambda L}\approx 4\%,
\end{align}
which is negligible in our calculation. When a power law decay $z^{-\alpha}$ with $\alpha>0$ is included in the asymptotic behavior, the finite volume effects are in general suppressed by an extra factor of $(\Lambda L)^{-\alpha}$ to the above estimation.

In 3D, analogous FV effects arise whenever $\vec{P} \neq (2\pi\vec{n}/L)$ with $\vec{n} \in \mathbb{Z}^3$.
They can again be seen to be exponentially suppressed by the Poisson summation formula, but their analytic form is more complicated than the 1D case.

\section{SRI'/MOM Renormalization}\label{app:renorm}

The renormalization of CG quasi-PDF operators reduces to the quark wave-function renormalization. It is suggested to calculate the renormalization factor in the static regularization-independent momentum-subtraction (SRI$'$/MOM) scheme~\cite{Gao:2026hix}. We follow the same procedure as Eq.~(A2) of Ref.~\cite{Gao:2026hix} by computing the equal-time quark propagator on the gauge-fixed lattice and comparing it with its tree-level lattice version in Eq.~(A3) of Ref.~\cite{Gao:2026hix}.
Then we subtract the {cubic lattice artifacts} using the parameterization in Eq.~(A5-A7) of Ref.~\cite{Gao:2026hix} motivated by lattice perturbation theory.
The subtracted SRI$'$ factor $Z_{\mathrm{SRI'}}^{-1/2}(|\vec{k}|) = c_1(\vec{k}^2) $ is then evolved to $\mu=2$ GeV and matched to $\msbar$ at one-loop order, as shown in Eq.~(A8-A9) of Ref.~\cite{Gao:2026hix}.

\begin{figure}[t]
    \centering
    \includegraphics[width=0.9\linewidth]{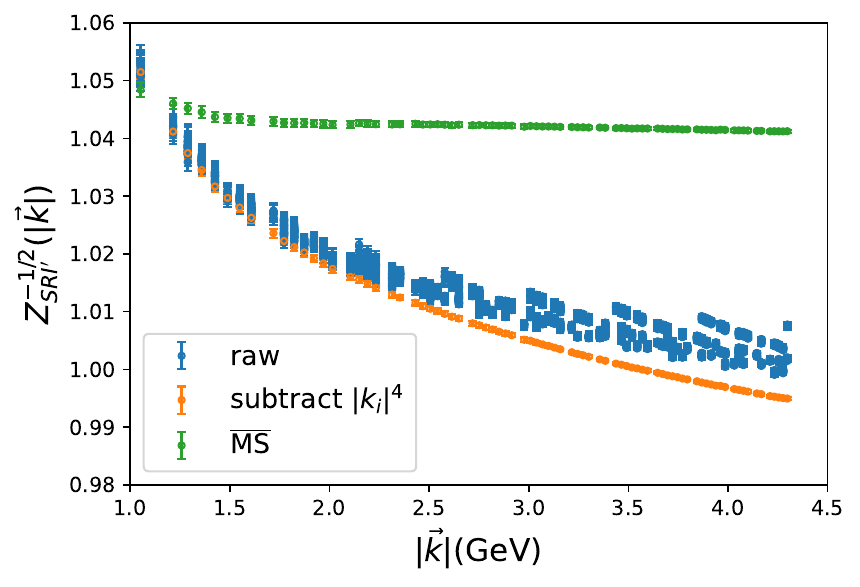}
    \caption{Comparison for the raw renormalization factor calculated in the SRI'/MOM scheme, after the subtraction of cubic lattice artifacts and the conversion to $\msbar$.}
    \label{fig:renorm_factor}
\end{figure}

The comparison between the raw $Z^{-1/2}_\mathrm{SRI'}(|\vec{k}|)$, the subtracted $Z^{-1/2}_\mathrm{SRI'}(|\vec{k}|)$, and the $\msbar$ results at $\mu=2$~GeV are shown in Fig.~\ref{fig:renorm_factor}. We determine the renormalization factor to be $Z^{\mathrm{\overline{\mathrm{MS}}}}=0.9215(4)$, almost independent of the quark momentum $|\vec{k}|$ across a large windowa.

\section{Discretization correction to $\msbar$ matching}\label{app:oa_corr}
Although we computed the multiplicative renormalization factor in the $\msbar$ scheme, the renormalized matrix elements in the $z/a\to0$ limit are different from the $\msbar$ perturbative calculation because the $z/a\to0$ limit does not commute with renormalization, introducing an $\mathcal{O}(a)$ discretization effect. The perturbative prediction~\cite{Gao:2023lny} suggests that
\begin{align}
    \lim_{z\to0}h^{\msbar}(z,P_z,\mu)= 1+\frac{\alpha_s(\mu)C_F}{4\pi}\left(1-\ln \frac{z^2\mu^2}{4e^{-2\gamma_E}}\right),
\end{align}
while on the lattice $h^R_{\rm lat}(z\to0)= Z_V/Z_C$ is a constant due to vector current conservation $h^B_{\rm lat}(0)/Z_V=1$. Since $Z_C$ is the renormalization constant in $\msbar$ converted from the lattice regulator $a$, $h^R_{\rm lat}(0)$ only differs in finite terms from $h^{\msbar}(a,P_z,\mu)$.
Thus the correction to the matching kernel can be estimated from the difference between $h^{\msbar}(a,P_z,\mu)$ and $h^{\msbar}(z,P_z,\mu)$ in the $|z|<a$ region, plus the $\mathcal{O}(\alpha_s)$ finite terms. The latter scales the value of $h^R_{\rm lat}(z)$ globally, thus modifying the matching kernel by a term proportional to $\delta(1-\xi)$,
\begin{align}\label{eq:improv}
    &\delta C^{(1)}(\xi)
    =\delta(1-\xi)[h^R_{\rm lat}(0)-h^{\msbar}(a)]+ \int_{-a}^a \frac{dzp_z}{2\pi}e^{i\xi zp_z} \nn
    &\qquad\times\int_{-1}^1 d\nu e^{-i\nu zp_z} \delta(1-\nu) [h^{\msbar}(a)-h^{\msbar}(z)]  \nn
    &= \delta(1-\xi)[h^R_{\rm lat}(0)-h^{\msbar}(a)]-\frac{\alpha_sC_F}{2\pi}\frac{{\rm Si}[(1-\xi) ap_z]}{\pi(1-\xi)},\nonumber
\end{align}
where $\xi=x/y$ and ${\rm Si}(t)=\int_0^t(\sin t')/t'dt'$.
\begin{figure}
    \centering
    \includegraphics[width=0.99\linewidth]{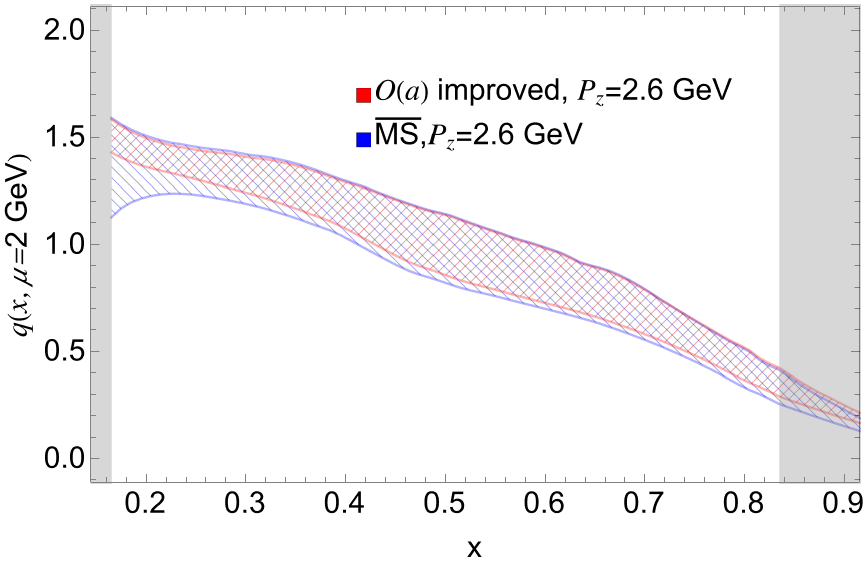}
    \caption{Comparison between the $\msbar$ PDFs before and after $\mathcal{O}(a)$ improvement with Eq.~\eqref{eq:improv}.}
    \label{fig:msbar_corr}
\end{figure}
$Z^{-1}_C$ contains the same $\ln a^2$ divergence as $h^{\msbar}(a)$ thus the first term vanishes in the continuum limit. The correction in momentum space is then $\mathcal{O}(a)$,
\begin{align}
    \delta C^{(1)}&(\xi)\xrightarrow{a\to0} -\frac{\alpha_sC_F}{2\pi^2}ap_z\,,
\end{align}
which has a well-defined continuum limit. For non-zero momentum, the perturbative result also depends on extra $\mathcal{O}(a^2)$ terms, which are negligible.
A comparison between the results with and without the $\mathcal{O}(a)$ improvement is shown in Fig.~\ref{fig:msbar_corr}. The correction in mid-$x$ region is much smaller than the overall uncertainties.

\section{$P_z$-evolution of the $k_T$-distribution}
\label{app:cs_evo}

The momentum evolution equation of the CG quasi-TMD is
\begin{align}
    {d \over d\ln P_z}\tilde{f}(x,b_T,\mu,P_z) = \gamma_q(b_T,\mu) + \delta\gamma_q(\mu,xP_z)\,,
\end{align}
where $\gamma_q(b_T,\mu)$ is the Collins-Soper kernel, and $\delta\gamma_q(\mu,xP_z)$ is known at one-loop order~\cite{Zhao:2023ptv}.

\begin{figure}[t]
    \centering
    \includegraphics[width=0.9\linewidth]{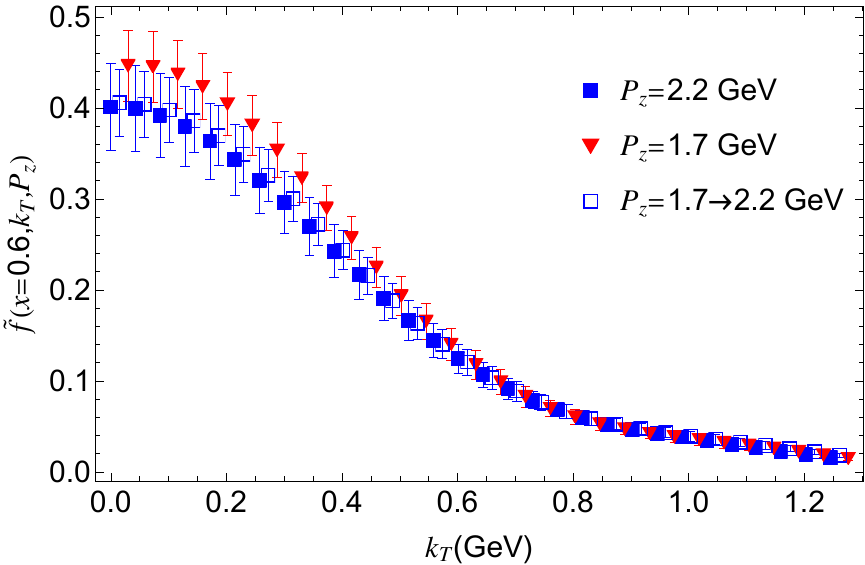}
    \caption{Comparison of the $k_T$-distributions at $x=0.6$ within a pion moving at $P_z=2.2$~GeV and $P_z=1.7$~GeV, and the latter evolved to $P_z=2.2$~GeV using the CS kernel calculated in Ref.~\cite{Avkhadiev:2024mgd}.}
    \label{fig:cs_evo}
\end{figure}

Therefore, we can calculate the quasi-TMD at an initial $P_z$ and evolve it to an arbitrary momentum, with the lattice CS kernel from Ref.~\cite{Avkhadiev:2024mgd}. The CS evolution factor is first applied to the lattice correlators in $\vec{b}_T$-space, then Fourier transformed to $\vec{k}_T$-space to extract the quasi-TMD. We compare the quasi-TMD calculated directly at $P_1^z=2.2$~GeV with that evolved from a quasi-TMD at $P_2^z=1.7$~GeV, and find consistent results within uncertainties across a broad range of $k_T$ for moderate $x$, as shown in Fig.~\ref{fig:cs_evo}. This demonstrates that the quasi-TMDs obtained from the momentum-space method satisfy the expected physical behavior.

\newpage

\begin{thebibliography}{78}%
\makeatletter
\providecommand \@ifxundefined [1]{%
 \@ifx{#1\undefined}
}%
\providecommand \@ifnum [1]{%
 \ifnum #1\expandafter \@firstoftwo
 \else \expandafter \@secondoftwo
 \fi
}%
\providecommand \@ifx [1]{%
 \ifx #1\expandafter \@firstoftwo
 \else \expandafter \@secondoftwo
 \fi
}%
\providecommand \natexlab [1]{#1}%
\providecommand \enquote  [1]{``#1''}%
\providecommand \bibnamefont  [1]{#1}%
\providecommand \bibfnamefont [1]{#1}%
\providecommand \citenamefont [1]{#1}%
\providecommand \href@noop [0]{\@secondoftwo}%
\providecommand \href [0]{\begingroup \@sanitize@url \@href}%
\providecommand \@href[1]{\@@startlink{#1}\@@href}%
\providecommand \@@href[1]{\endgroup#1\@@endlink}%
\providecommand \@sanitize@url [0]{\catcode `\\12\catcode `\$12\catcode
  `\&12\catcode `\#12\catcode `\^12\catcode `\_12\catcode `\%12\relax}%
\providecommand \@@startlink[1]{}%
\providecommand \@@endlink[0]{}%
\providecommand \url  [0]{\begingroup\@sanitize@url \@url }%
\providecommand \@url [1]{\endgroup\@href {#1}{\urlprefix }}%
\providecommand \urlprefix  [0]{URL }%
\providecommand \Eprint [0]{\href }%
\providecommand \doibase [0]{http://dx.doi.org/}%
\providecommand \selectlanguage [0]{\@gobble}%
\providecommand \bibinfo  [0]{\@secondoftwo}%
\providecommand \bibfield  [0]{\@secondoftwo}%
\providecommand \translation [1]{[#1]}%
\providecommand \BibitemOpen [0]{}%
\providecommand \bibitemStop [0]{}%
\providecommand \bibitemNoStop [0]{.\EOS\space}%
\providecommand \EOS [0]{\spacefactor3000\relax}%
\providecommand \BibitemShut  [1]{\csname bibitem#1\endcsname}%
\let\auto@bib@innerbib\@empty
\bibitem [{\citenamefont {Dudek}\ \emph {et~al.}(2012)\citenamefont {Dudek}
  \emph {et~al.}}]{Dudek:2012vr}%
  \BibitemOpen
  \bibfield  {author} {\bibinfo {author} {\bibfnamefont {J.}~\bibnamefont
  {Dudek}} \emph {et~al.},\ }\href {\doibase 10.1140/epja/i2012-12187-1}
  {\bibfield  {journal} {\bibinfo  {journal} {Eur. Phys. J. A}\ }\textbf
  {\bibinfo {volume} {48}},\ \bibinfo {pages} {187} (\bibinfo {year} {2012})},\
  \Eprint {http://arxiv.org/abs/1208.1244} {arXiv:1208.1244 [hep-ex]}
  \BibitemShut {NoStop}%
\bibitem [{\citenamefont {Accardi}\ \emph {et~al.}(2016)\citenamefont {Accardi}
  \emph {et~al.}}]{Accardi:2012qut}%
  \BibitemOpen
  \bibfield  {author} {\bibinfo {author} {\bibfnamefont {A.}~\bibnamefont
  {Accardi}} \emph {et~al.},\ }\href {\doibase 10.1140/epja/i2016-16268-9}
  {\bibfield  {journal} {\bibinfo  {journal} {Eur. Phys. J. A}\ }\textbf
  {\bibinfo {volume} {52}},\ \bibinfo {pages} {268} (\bibinfo {year} {2016})},\
  \Eprint {http://arxiv.org/abs/1212.1701} {arXiv:1212.1701 [nucl-ex]}
  \BibitemShut {NoStop}%
\bibitem [{\citenamefont {Abdul~Khalek}\ \emph {et~al.}(2022)\citenamefont
  {Abdul~Khalek} \emph {et~al.}}]{AbdulKhalek:2021gbh}%
  \BibitemOpen
  \bibfield  {author} {\bibinfo {author} {\bibfnamefont {R.}~\bibnamefont
  {Abdul~Khalek}} \emph {et~al.},\ }\href {\doibase
  10.1016/j.nuclphysa.2022.122447} {\bibfield  {journal} {\bibinfo  {journal}
  {Nucl. Phys. A}\ }\textbf {\bibinfo {volume} {1026}},\ \bibinfo {pages}
  {122447} (\bibinfo {year} {2022})},\ \Eprint
  {http://arxiv.org/abs/2103.05419} {arXiv:2103.05419 [physics.ins-det]}
  \BibitemShut {NoStop}%
\bibitem [{\citenamefont {Apollinari}\ \emph {et~al.}(2015)\citenamefont
  {Apollinari}, \citenamefont {Br{\"u}ning}, \citenamefont {Nakamoto},\ and\
  \citenamefont {Rossi}}]{Apollinari:2015wtw}%
  \BibitemOpen
  \bibfield  {author} {\bibinfo {author} {\bibfnamefont {G.}~\bibnamefont
  {Apollinari}}, \bibinfo {author} {\bibfnamefont {O.}~\bibnamefont
  {Br{\"u}ning}}, \bibinfo {author} {\bibfnamefont {T.}~\bibnamefont
  {Nakamoto}}, \ and\ \bibinfo {author} {\bibfnamefont {L.}~\bibnamefont
  {Rossi}},\ }\href {\doibase 10.5170/CERN-2015-005.1} {\bibfield  {journal}
  {\bibinfo  {journal} {CERN Yellow Rep.}\ ,\ \bibinfo {pages} {1}} (\bibinfo
  {year} {2015})},\ \Eprint {http://arxiv.org/abs/1705.08830} {arXiv:1705.08830
  [physics.acc-ph]} \BibitemShut {NoStop}%
\bibitem [{\citenamefont {Constantinou}\ \emph {et~al.}(2021)\citenamefont
  {Constantinou} \emph {et~al.}}]{Constantinou:2020hdm}%
  \BibitemOpen
  \bibfield  {author} {\bibinfo {author} {\bibfnamefont {M.}~\bibnamefont
  {Constantinou}} \emph {et~al.},\ }\href {\doibase 10.1016/j.ppnp.2021.103908}
  {\bibfield  {journal} {\bibinfo  {journal} {Prog. Part. Nucl. Phys.}\
  }\textbf {\bibinfo {volume} {121}},\ \bibinfo {pages} {103908} (\bibinfo
  {year} {2021})},\ \Eprint {http://arxiv.org/abs/2006.08636} {arXiv:2006.08636
  [hep-ph]} \BibitemShut {NoStop}%
\bibitem [{\citenamefont {Kronfeld}\ and\ \citenamefont
  {Photiadis}(1985)}]{Kronfeld:1984zv}%
  \BibitemOpen
  \bibfield  {author} {\bibinfo {author} {\bibfnamefont {A.~S.}\ \bibnamefont
  {Kronfeld}}\ and\ \bibinfo {author} {\bibfnamefont {D.~M.}\ \bibnamefont
  {Photiadis}},\ }\href {\doibase 10.1103/PhysRevD.31.2939} {\bibfield
  {journal} {\bibinfo  {journal} {Phys. Rev. D}\ }\textbf {\bibinfo {volume}
  {31}},\ \bibinfo {pages} {2939} (\bibinfo {year} {1985})}\BibitemShut
  {NoStop}%
\bibitem [{\citenamefont {Martinelli}\ and\ \citenamefont
  {Sachrajda}(1987)}]{Martinelli:1987si}%
  \BibitemOpen
  \bibfield  {author} {\bibinfo {author} {\bibfnamefont {G.}~\bibnamefont
  {Martinelli}}\ and\ \bibinfo {author} {\bibfnamefont {C.~T.}\ \bibnamefont
  {Sachrajda}},\ }\href {\doibase 10.1016/0370-2693(87)90858-6} {\bibfield
  {journal} {\bibinfo  {journal} {Phys. Lett. B}\ }\textbf {\bibinfo {volume}
  {190}},\ \bibinfo {pages} {151} (\bibinfo {year} {1987})}\BibitemShut
  {NoStop}%
\bibitem [{\citenamefont {Davoudi}\ and\ \citenamefont
  {Savage}(2012)}]{Davoudi:2012ya}%
  \BibitemOpen
  \bibfield  {author} {\bibinfo {author} {\bibfnamefont {Z.}~\bibnamefont
  {Davoudi}}\ and\ \bibinfo {author} {\bibfnamefont {M.~J.}\ \bibnamefont
  {Savage}},\ }\href {\doibase 10.1103/PhysRevD.86.054505} {\bibfield
  {journal} {\bibinfo  {journal} {Phys. Rev. D}\ }\textbf {\bibinfo {volume}
  {86}},\ \bibinfo {pages} {054505} (\bibinfo {year} {2012})},\ \Eprint
  {http://arxiv.org/abs/1204.4146} {arXiv:1204.4146 [hep-lat]} \BibitemShut
  {NoStop}%
\bibitem [{\citenamefont {Shindler}(2024)}]{Shindler:2023xpd}%
  \BibitemOpen
  \bibfield  {author} {\bibinfo {author} {\bibfnamefont {A.}~\bibnamefont
  {Shindler}},\ }\href {\doibase 10.1103/PhysRevD.110.L051503} {\bibfield
  {journal} {\bibinfo  {journal} {Phys. Rev. D}\ }\textbf {\bibinfo {volume}
  {110}},\ \bibinfo {pages} {L051503} (\bibinfo {year} {2024})},\ \Eprint
  {http://arxiv.org/abs/2311.18704} {arXiv:2311.18704 [hep-lat]} \BibitemShut
  {NoStop}%
\bibitem [{\citenamefont {Braun}\ and\ \citenamefont
  {M{\"u}ller}(2008)}]{Braun:2007wv}%
  \BibitemOpen
  \bibfield  {author} {\bibinfo {author} {\bibfnamefont {V.}~\bibnamefont
  {Braun}}\ and\ \bibinfo {author} {\bibfnamefont {D.}~\bibnamefont
  {M{\"u}ller}},\ }\href {\doibase 10.1140/epjc/s10052-008-0608-4} {\bibfield
  {journal} {\bibinfo  {journal} {Eur. Phys. J. C}\ }\textbf {\bibinfo {volume}
  {55}},\ \bibinfo {pages} {349} (\bibinfo {year} {2008})},\ \Eprint
  {http://arxiv.org/abs/0709.1348} {arXiv:0709.1348 [hep-ph]} \BibitemShut
  {NoStop}%
\bibitem [{\citenamefont {Radyushkin}(2017)}]{Radyushkin:2017cyf}%
  \BibitemOpen
  \bibfield  {author} {\bibinfo {author} {\bibfnamefont {A.~V.}\ \bibnamefont
  {Radyushkin}},\ }\href {\doibase 10.1103/PhysRevD.96.034025} {\bibfield
  {journal} {\bibinfo  {journal} {Phys. Rev. D}\ }\textbf {\bibinfo {volume}
  {96}},\ \bibinfo {pages} {034025} (\bibinfo {year} {2017})},\ \Eprint
  {http://arxiv.org/abs/1705.01488} {arXiv:1705.01488 [hep-ph]} \BibitemShut
  {NoStop}%
\bibitem [{\citenamefont {Ma}\ and\ \citenamefont {Qiu}(2018)}]{Ma:2017pxb}%
  \BibitemOpen
  \bibfield  {author} {\bibinfo {author} {\bibfnamefont {Y.-Q.}\ \bibnamefont
  {Ma}}\ and\ \bibinfo {author} {\bibfnamefont {J.-W.}\ \bibnamefont {Qiu}},\
  }\href {\doibase 10.1103/PhysRevLett.120.022003} {\bibfield  {journal}
  {\bibinfo  {journal} {Phys. Rev. Lett.}\ }\textbf {\bibinfo {volume} {120}},\
  \bibinfo {pages} {022003} (\bibinfo {year} {2018})},\ \Eprint
  {http://arxiv.org/abs/1709.03018} {arXiv:1709.03018 [hep-ph]} \BibitemShut
  {NoStop}%
\bibitem [{\citenamefont {Liu}\ and\ \citenamefont {Dong}(1994)}]{Liu:1993cv}%
  \BibitemOpen
  \bibfield  {author} {\bibinfo {author} {\bibfnamefont {K.-F.}\ \bibnamefont
  {Liu}}\ and\ \bibinfo {author} {\bibfnamefont {S.-J.}\ \bibnamefont {Dong}},\
  }\href {\doibase 10.1103/PhysRevLett.72.1790} {\bibfield  {journal} {\bibinfo
   {journal} {Phys. Rev. Lett.}\ }\textbf {\bibinfo {volume} {72}},\ \bibinfo
  {pages} {1790} (\bibinfo {year} {1994})},\ \Eprint
  {http://arxiv.org/abs/hep-ph/9306299} {arXiv:hep-ph/9306299} \BibitemShut
  {NoStop}%
\bibitem [{\citenamefont {Detmold}\ and\ \citenamefont
  {Lin}(2006)}]{Detmold:2005gg}%
  \BibitemOpen
  \bibfield  {author} {\bibinfo {author} {\bibfnamefont {W.}~\bibnamefont
  {Detmold}}\ and\ \bibinfo {author} {\bibfnamefont {C.~J.~D.}\ \bibnamefont
  {Lin}},\ }\href {\doibase 10.1103/PhysRevD.73.014501} {\bibfield  {journal}
  {\bibinfo  {journal} {Phys. Rev. D}\ }\textbf {\bibinfo {volume} {73}},\
  \bibinfo {pages} {014501} (\bibinfo {year} {2006})},\ \Eprint
  {http://arxiv.org/abs/hep-lat/0507007} {arXiv:hep-lat/0507007} \BibitemShut
  {NoStop}%
\bibitem [{\citenamefont {Detmold}\ \emph {et~al.}(2021)\citenamefont
  {Detmold}, \citenamefont {Grebe}, \citenamefont {Kanamori}, \citenamefont
  {Lin}, \citenamefont {Perry},\ and\ \citenamefont {Zhao}}]{Detmold:2021uru}%
  \BibitemOpen
  \bibfield  {author} {\bibinfo {author} {\bibfnamefont {W.}~\bibnamefont
  {Detmold}}, \bibinfo {author} {\bibfnamefont {A.~V.}\ \bibnamefont {Grebe}},
  \bibinfo {author} {\bibfnamefont {I.}~\bibnamefont {Kanamori}}, \bibinfo
  {author} {\bibfnamefont {C.~J.~D.}\ \bibnamefont {Lin}}, \bibinfo {author}
  {\bibfnamefont {R.~J.}\ \bibnamefont {Perry}}, \ and\ \bibinfo {author}
  {\bibfnamefont {Y.}~\bibnamefont {Zhao}} (\bibinfo {collaboration} {HOPE}),\
  }\href {\doibase 10.1103/PhysRevD.104.074511} {\bibfield  {journal} {\bibinfo
   {journal} {Phys. Rev. D}\ }\textbf {\bibinfo {volume} {104}},\ \bibinfo
  {pages} {074511} (\bibinfo {year} {2021})},\ \Eprint
  {http://arxiv.org/abs/2103.09529} {arXiv:2103.09529 [hep-lat]} \BibitemShut
  {NoStop}%
\bibitem [{\citenamefont {Chambers}\ \emph {et~al.}(2017)\citenamefont
  {Chambers}, \citenamefont {Horsley}, \citenamefont {Nakamura}, \citenamefont
  {Perlt}, \citenamefont {Rakow}, \citenamefont {Schierholz}, \citenamefont
  {Schiller}, \citenamefont {Somfleth}, \citenamefont {Young},\ and\
  \citenamefont {Zanotti}}]{Chambers:2017dov}%
  \BibitemOpen
  \bibfield  {author} {\bibinfo {author} {\bibfnamefont {A.~J.}\ \bibnamefont
  {Chambers}}, \bibinfo {author} {\bibfnamefont {R.}~\bibnamefont {Horsley}},
  \bibinfo {author} {\bibfnamefont {Y.}~\bibnamefont {Nakamura}}, \bibinfo
  {author} {\bibfnamefont {H.}~\bibnamefont {Perlt}}, \bibinfo {author}
  {\bibfnamefont {P.~E.~L.}\ \bibnamefont {Rakow}}, \bibinfo {author}
  {\bibfnamefont {G.}~\bibnamefont {Schierholz}}, \bibinfo {author}
  {\bibfnamefont {A.}~\bibnamefont {Schiller}}, \bibinfo {author}
  {\bibfnamefont {K.}~\bibnamefont {Somfleth}}, \bibinfo {author}
  {\bibfnamefont {R.~D.}\ \bibnamefont {Young}}, \ and\ \bibinfo {author}
  {\bibfnamefont {J.~M.}\ \bibnamefont {Zanotti}},\ }\href {\doibase
  10.1103/PhysRevLett.118.242001} {\bibfield  {journal} {\bibinfo  {journal}
  {Phys. Rev. Lett.}\ }\textbf {\bibinfo {volume} {118}},\ \bibinfo {pages}
  {242001} (\bibinfo {year} {2017})},\ \Eprint
  {http://arxiv.org/abs/1703.01153} {arXiv:1703.01153 [hep-lat]} \BibitemShut
  {NoStop}%
\bibitem [{\citenamefont {Ji}(2013)}]{Ji:2013dva}%
  \BibitemOpen
  \bibfield  {author} {\bibinfo {author} {\bibfnamefont {X.}~\bibnamefont
  {Ji}},\ }\href {\doibase 10.1103/PhysRevLett.110.262002} {\bibfield
  {journal} {\bibinfo  {journal} {Phys. Rev. Lett.}\ }\textbf {\bibinfo
  {volume} {110}},\ \bibinfo {pages} {262002} (\bibinfo {year} {2013})},\
  \Eprint {http://arxiv.org/abs/1305.1539} {arXiv:1305.1539 [hep-ph]}
  \BibitemShut {NoStop}%
\bibitem [{\citenamefont {Ji}(2014)}]{Ji:2014gla}%
  \BibitemOpen
  \bibfield  {author} {\bibinfo {author} {\bibfnamefont {X.}~\bibnamefont
  {Ji}},\ }\href {\doibase 10.1007/s11433-014-5492-3} {\bibfield  {journal}
  {\bibinfo  {journal} {Sci. China Phys. Mech. Astron.}\ }\textbf {\bibinfo
  {volume} {57}},\ \bibinfo {pages} {1407} (\bibinfo {year} {2014})},\ \Eprint
  {http://arxiv.org/abs/1404.6680} {arXiv:1404.6680 [hep-ph]} \BibitemShut
  {NoStop}%
\bibitem [{\citenamefont {Ji}\ \emph {et~al.}(2021{\natexlab{a}})\citenamefont
  {Ji}, \citenamefont {Liu}, \citenamefont {Liu}, \citenamefont {Zhang},\ and\
  \citenamefont {Zhao}}]{Ji:2020ect}%
  \BibitemOpen
  \bibfield  {author} {\bibinfo {author} {\bibfnamefont {X.}~\bibnamefont
  {Ji}}, \bibinfo {author} {\bibfnamefont {Y.-S.}\ \bibnamefont {Liu}},
  \bibinfo {author} {\bibfnamefont {Y.}~\bibnamefont {Liu}}, \bibinfo {author}
  {\bibfnamefont {J.-H.}\ \bibnamefont {Zhang}}, \ and\ \bibinfo {author}
  {\bibfnamefont {Y.}~\bibnamefont {Zhao}},\ }\href {\doibase
  10.1103/RevModPhys.93.035005} {\bibfield  {journal} {\bibinfo  {journal}
  {Rev. Mod. Phys.}\ }\textbf {\bibinfo {volume} {93}},\ \bibinfo {pages}
  {035005} (\bibinfo {year} {2021}{\natexlab{a}})},\ \Eprint
  {http://arxiv.org/abs/2004.03543} {arXiv:2004.03543 [hep-ph]} \BibitemShut
  {NoStop}%
\bibitem [{\citenamefont {Ji}(2020)}]{Ji:2020byp}%
  \BibitemOpen
  \bibfield  {author} {\bibinfo {author} {\bibfnamefont {X.}~\bibnamefont
  {Ji}},\ }\href@noop {} {\  (\bibinfo {year} {2020})},\ \Eprint
  {http://arxiv.org/abs/2007.06613} {arXiv:2007.06613 [hep-ph]} \BibitemShut
  {NoStop}%
\bibitem [{\citenamefont {Ji}(2024)}]{Ji:2024oka}%
  \BibitemOpen
  \bibfield  {author} {\bibinfo {author} {\bibfnamefont {X.}~\bibnamefont
  {Ji}},\ }\href {\doibase 10.1016/j.nuclphysb.2024.116670} {\bibfield
  {journal} {\bibinfo  {journal} {Nucl. Phys. B}\ }\textbf {\bibinfo {volume}
  {1007}},\ \bibinfo {pages} {116670} (\bibinfo {year} {2024})},\ \Eprint
  {http://arxiv.org/abs/2408.03378} {arXiv:2408.03378 [hep-ph]} \BibitemShut
  {NoStop}%
\bibitem [{\citenamefont {Lin}(2025)}]{Lin:2025hka}%
  \BibitemOpen
  \bibfield  {author} {\bibinfo {author} {\bibfnamefont {H.-W.}\ \bibnamefont
  {Lin}},\ }\href {\doibase 10.1016/j.ppnp.2025.104177} {\bibfield  {journal}
  {\bibinfo  {journal} {Prog. Part. Nucl. Phys.}\ }\textbf {\bibinfo {volume}
  {144}},\ \bibinfo {pages} {104177} (\bibinfo {year} {2025})},\ \Eprint
  {http://arxiv.org/abs/2506.05025} {arXiv:2506.05025 [hep-lat]} \BibitemShut
  {NoStop}%
\bibitem [{\citenamefont {Zhao}(2025)}]{Zhao:2025oto}%
  \BibitemOpen
  \bibfield  {author} {\bibinfo {author} {\bibfnamefont {Y.}~\bibnamefont
  {Zhao}},\ }\href@noop {} {\  (\bibinfo {year} {2025})},\ \Eprint
  {http://arxiv.org/abs/2509.00247} {arXiv:2509.00247 [hep-lat]} \BibitemShut
  {NoStop}%
\bibitem [{INT(2024)}]{INT24-88W}%
  \BibitemOpen
  \href@noop {} {\enquote {\bibinfo {title} {Inverse problems and uncertainty
  quantification in nuclear physics},}\ }\bibinfo {howpublished}
  {\url{https://www.int.washington.edu/programs-and-workshops/24-88w}}
  (\bibinfo {year} {2024}),\ \bibinfo {note} {workshop held at Institute for
  Nuclear Theory, July 8--12, 2024; Event ID: INT-24-88W}\BibitemShut {NoStop}%
\bibitem [{PDF(2024)}]{PDFLattice2024}%
  \BibitemOpen
  \href@noop {} {\enquote {\bibinfo {title} {Parton distributions and lattice
  calculations (pdflattice 2024)},}\ }\bibinfo {howpublished}
  {\url{https://indico.cern.ch/event/1434067/}} (\bibinfo {year} {2024}),\
  \bibinfo {note} {workshop held at Jefferson Lab, Nov 18--20,
  2024}\BibitemShut {NoStop}%
\bibitem [{\citenamefont {Del~Debbio}\ \emph {et~al.}(2024)\citenamefont
  {Del~Debbio}, \citenamefont {Lupo}, \citenamefont {Panero},\ and\
  \citenamefont {Tantalo}}]{DelDebbio:2024sfa}%
  \BibitemOpen
  \bibfield  {author} {\bibinfo {author} {\bibfnamefont {L.}~\bibnamefont
  {Del~Debbio}}, \bibinfo {author} {\bibfnamefont {A.}~\bibnamefont {Lupo}},
  \bibinfo {author} {\bibfnamefont {M.}~\bibnamefont {Panero}}, \ and\ \bibinfo
  {author} {\bibfnamefont {N.}~\bibnamefont {Tantalo}},\ }in\ \href@noop {}
  {\emph {\bibinfo {booktitle} {{EuroPLEx Final Conference}}}}\ (\bibinfo
  {year} {2024})\ \Eprint {http://arxiv.org/abs/2410.09944} {arXiv:2410.09944
  [hep-lat]} \BibitemShut {NoStop}%
\bibitem [{\citenamefont {Dutrieux}\ \emph
  {et~al.}(2025{\natexlab{a}})\citenamefont {Dutrieux}, \citenamefont {Karpie},
  \citenamefont {Monahan}, \citenamefont {Orginos}, \citenamefont {Radyushkin},
  \citenamefont {Richards},\ and\ \citenamefont
  {Zafeiropoulos}}]{Dutrieux:2025axb}%
  \BibitemOpen
  \bibfield  {author} {\bibinfo {author} {\bibfnamefont {H.}~\bibnamefont
  {Dutrieux}}, \bibinfo {author} {\bibfnamefont {J.}~\bibnamefont {Karpie}},
  \bibinfo {author} {\bibfnamefont {C.~J.}\ \bibnamefont {Monahan}}, \bibinfo
  {author} {\bibfnamefont {K.}~\bibnamefont {Orginos}}, \bibinfo {author}
  {\bibfnamefont {A.}~\bibnamefont {Radyushkin}}, \bibinfo {author}
  {\bibfnamefont {D.}~\bibnamefont {Richards}}, \ and\ \bibinfo {author}
  {\bibfnamefont {S.}~\bibnamefont {Zafeiropoulos}},\ }\href@noop {} {\
  (\bibinfo {year} {2025}{\natexlab{a}})},\ \Eprint
  {http://arxiv.org/abs/2506.24037} {arXiv:2506.24037 [hep-lat]} \BibitemShut
  {NoStop}%
\bibitem [{\citenamefont {Chen}\ \emph {et~al.}(2026)\citenamefont {Chen} \emph
  {et~al.}}]{Chen:2025cxr}%
  \BibitemOpen
  \bibfield  {author} {\bibinfo {author} {\bibfnamefont {J.-W.}\ \bibnamefont
  {Chen}} \emph {et~al.},\ }\href {\doibase 10.1103/fflw-qpcc} {\bibfield
  {journal} {\bibinfo  {journal} {Phys. Rev. D}\ }\textbf {\bibinfo {volume}
  {113}},\ \bibinfo {pages} {014509} (\bibinfo {year} {2026})},\ \Eprint
  {http://arxiv.org/abs/2505.14619} {arXiv:2505.14619 [hep-lat]} \BibitemShut
  {NoStop}%
\bibitem [{\citenamefont {Xiong}\ \emph {et~al.}(2025)\citenamefont {Xiong},
  \citenamefont {Hua}, \citenamefont {Ling}, \citenamefont {Wei}, \citenamefont
  {Yu}, \citenamefont {Zhang},\ and\ \citenamefont {Zheng}}]{Xiong:2025obq}%
  \BibitemOpen
  \bibfield  {author} {\bibinfo {author} {\bibfnamefont {A.-S.}\ \bibnamefont
  {Xiong}}, \bibinfo {author} {\bibfnamefont {J.}~\bibnamefont {Hua}}, \bibinfo
  {author} {\bibfnamefont {Y.-F.}\ \bibnamefont {Ling}}, \bibinfo {author}
  {\bibfnamefont {T.}~\bibnamefont {Wei}}, \bibinfo {author} {\bibfnamefont
  {F.-S.}\ \bibnamefont {Yu}}, \bibinfo {author} {\bibfnamefont {Q.-A.}\
  \bibnamefont {Zhang}}, \ and\ \bibinfo {author} {\bibfnamefont
  {Y.}~\bibnamefont {Zheng}},\ }\href {\doibase
  10.1140/epjc/s10052-025-15130-9} {\bibfield  {journal} {\bibinfo  {journal}
  {Eur. Phys. J. C}\ }\textbf {\bibinfo {volume} {85}},\ \bibinfo {pages}
  {1409} (\bibinfo {year} {2025})},\ \Eprint {http://arxiv.org/abs/2506.16689}
  {arXiv:2506.16689 [hep-lat]} \BibitemShut {NoStop}%
\bibitem [{\citenamefont {Chu}\ \emph {et~al.}(2025)\citenamefont {Chu},
  \citenamefont {Cichy}, \citenamefont {Constantinou}, \citenamefont
  {Sznajder},\ and\ \citenamefont {Wagner}}]{Chu:2025jsi}%
  \BibitemOpen
  \bibfield  {author} {\bibinfo {author} {\bibfnamefont {M.-H.}\ \bibnamefont
  {Chu}}, \bibinfo {author} {\bibfnamefont {K.}~\bibnamefont {Cichy}}, \bibinfo
  {author} {\bibfnamefont {M.}~\bibnamefont {Constantinou}}, \bibinfo {author}
  {\bibfnamefont {P.}~\bibnamefont {Sznajder}}, \ and\ \bibinfo {author}
  {\bibfnamefont {J.}~\bibnamefont {Wagner}},\ }\href@noop {} {\  (\bibinfo
  {year} {2025})},\ \Eprint {http://arxiv.org/abs/2509.15931} {arXiv:2509.15931
  [hep-lat]} \BibitemShut {NoStop}%
\bibitem [{\citenamefont {Dutrieux}\ \emph
  {et~al.}(2025{\natexlab{b}})\citenamefont {Dutrieux}, \citenamefont {Karpie},
  \citenamefont {Monahan}, \citenamefont {Orginos}, \citenamefont {Radyushkin},
  \citenamefont {Richards},\ and\ \citenamefont
  {Zafeiropoulos}}]{Dutrieux:2025jed}%
  \BibitemOpen
  \bibfield  {author} {\bibinfo {author} {\bibfnamefont {H.}~\bibnamefont
  {Dutrieux}}, \bibinfo {author} {\bibfnamefont {J.}~\bibnamefont {Karpie}},
  \bibinfo {author} {\bibfnamefont {C.~J.}\ \bibnamefont {Monahan}}, \bibinfo
  {author} {\bibfnamefont {K.}~\bibnamefont {Orginos}}, \bibinfo {author}
  {\bibfnamefont {A.}~\bibnamefont {Radyushkin}}, \bibinfo {author}
  {\bibfnamefont {D.}~\bibnamefont {Richards}}, \ and\ \bibinfo {author}
  {\bibfnamefont {S.}~\bibnamefont {Zafeiropoulos}},\ }\href@noop {} {\
  (\bibinfo {year} {2025}{\natexlab{b}})},\ \Eprint
  {http://arxiv.org/abs/2504.17706} {arXiv:2504.17706 [hep-lat]} \BibitemShut
  {NoStop}%
\bibitem [{\citenamefont {Ling}\ \emph {et~al.}(2025)\citenamefont {Ling},
  \citenamefont {Chu}, \citenamefont {Liang}, \citenamefont {Hua},
  \citenamefont {Xiong},\ and\ \citenamefont {Zhang}}]{Ling:2025olz}%
  \BibitemOpen
  \bibfield  {author} {\bibinfo {author} {\bibfnamefont {Y.-F.}\ \bibnamefont
  {Ling}}, \bibinfo {author} {\bibfnamefont {M.-H.}\ \bibnamefont {Chu}},
  \bibinfo {author} {\bibfnamefont {J.}~\bibnamefont {Liang}}, \bibinfo
  {author} {\bibfnamefont {J.}~\bibnamefont {Hua}}, \bibinfo {author}
  {\bibfnamefont {A.-S.}\ \bibnamefont {Xiong}}, \ and\ \bibinfo {author}
  {\bibfnamefont {Q.-A.}\ \bibnamefont {Zhang}},\ }\href@noop {} {\  (\bibinfo
  {year} {2025})},\ \Eprint {http://arxiv.org/abs/2511.03593} {arXiv:2511.03593
  [hep-lat]} \BibitemShut {NoStop}%
\bibitem [{\citenamefont {Rothkopf}(2026)}]{Rothkopf:2026wdj}%
  \BibitemOpen
  \bibfield  {author} {\bibinfo {author} {\bibfnamefont {A.}~\bibnamefont
  {Rothkopf}}\ }(\bibinfo {year} {2026})\ \Eprint
  {http://arxiv.org/abs/2604.01996} {arXiv:2604.01996 [hep-lat]} \BibitemShut
  {NoStop}%
\bibitem [{\citenamefont {Ji}\ \emph {et~al.}(2021{\natexlab{b}})\citenamefont
  {Ji}, \citenamefont {Liu}, \citenamefont {Sch\"afer}, \citenamefont {Wang},
  \citenamefont {Yang}, \citenamefont {Zhang},\ and\ \citenamefont
  {Zhao}}]{Ji:2020brr}%
  \BibitemOpen
  \bibfield  {author} {\bibinfo {author} {\bibfnamefont {X.}~\bibnamefont
  {Ji}}, \bibinfo {author} {\bibfnamefont {Y.}~\bibnamefont {Liu}}, \bibinfo
  {author} {\bibfnamefont {A.}~\bibnamefont {Sch\"afer}}, \bibinfo {author}
  {\bibfnamefont {W.}~\bibnamefont {Wang}}, \bibinfo {author} {\bibfnamefont
  {Y.-B.}\ \bibnamefont {Yang}}, \bibinfo {author} {\bibfnamefont {J.-H.}\
  \bibnamefont {Zhang}}, \ and\ \bibinfo {author} {\bibfnamefont
  {Y.}~\bibnamefont {Zhao}},\ }\href {\doibase 10.1016/j.nuclphysb.2021.115311}
  {\bibfield  {journal} {\bibinfo  {journal} {Nucl. Phys. B}\ }\textbf
  {\bibinfo {volume} {964}},\ \bibinfo {pages} {115311} (\bibinfo {year}
  {2021}{\natexlab{b}})},\ \Eprint {http://arxiv.org/abs/2008.03886}
  {arXiv:2008.03886 [hep-ph]} \BibitemShut {NoStop}%
\bibitem [{\citenamefont {Gao}\ \emph {et~al.}(2022)\citenamefont {Gao},
  \citenamefont {Hanlon}, \citenamefont {Mukherjee}, \citenamefont {Petreczky},
  \citenamefont {Scior}, \citenamefont {Syritsyn},\ and\ \citenamefont
  {Zhao}}]{Gao:2021dbh}%
  \BibitemOpen
  \bibfield  {author} {\bibinfo {author} {\bibfnamefont {X.}~\bibnamefont
  {Gao}}, \bibinfo {author} {\bibfnamefont {A.~D.}\ \bibnamefont {Hanlon}},
  \bibinfo {author} {\bibfnamefont {S.}~\bibnamefont {Mukherjee}}, \bibinfo
  {author} {\bibfnamefont {P.}~\bibnamefont {Petreczky}}, \bibinfo {author}
  {\bibfnamefont {P.}~\bibnamefont {Scior}}, \bibinfo {author} {\bibfnamefont
  {S.}~\bibnamefont {Syritsyn}}, \ and\ \bibinfo {author} {\bibfnamefont
  {Y.}~\bibnamefont {Zhao}},\ }\href {\doibase 10.1103/PhysRevLett.128.142003}
  {\bibfield  {journal} {\bibinfo  {journal} {Phys. Rev. Lett.}\ }\textbf
  {\bibinfo {volume} {128}},\ \bibinfo {pages} {142003} (\bibinfo {year}
  {2022})},\ \Eprint {http://arxiv.org/abs/2112.02208} {arXiv:2112.02208
  [hep-lat]} \BibitemShut {NoStop}%
\bibitem [{\citenamefont {Ji}\ \emph {et~al.}(2026)\citenamefont {Ji},
  \citenamefont {Liu},\ and\ \citenamefont {Su}}]{Ji:2026vir}%
  \BibitemOpen
  \bibfield  {author} {\bibinfo {author} {\bibfnamefont {X.}~\bibnamefont
  {Ji}}, \bibinfo {author} {\bibfnamefont {Y.}~\bibnamefont {Liu}}, \ and\
  \bibinfo {author} {\bibfnamefont {Y.}~\bibnamefont {Su}},\ }\href@noop {} {\
  (\bibinfo {year} {2026})},\ \Eprint {http://arxiv.org/abs/2601.12189}
  {arXiv:2601.12189 [hep-lat]} \BibitemShut {NoStop}%
\bibitem [{\citenamefont {Dotsenko}\ and\ \citenamefont
  {Vergeles}(1980)}]{Dotsenko:1979wb}%
  \BibitemOpen
  \bibfield  {author} {\bibinfo {author} {\bibfnamefont {V.~S.}\ \bibnamefont
  {Dotsenko}}\ and\ \bibinfo {author} {\bibfnamefont {S.~N.}\ \bibnamefont
  {Vergeles}},\ }\href {\doibase 10.1016/0550-3213(80)90103-0} {\bibfield
  {journal} {\bibinfo  {journal} {Nucl. Phys. B}\ }\textbf {\bibinfo {volume}
  {169}},\ \bibinfo {pages} {527} (\bibinfo {year} {1980})}\BibitemShut
  {NoStop}%
\bibitem [{\citenamefont {Craigie}\ and\ \citenamefont
  {Dorn}(1981)}]{Craigie:1980qs}%
  \BibitemOpen
  \bibfield  {author} {\bibinfo {author} {\bibfnamefont {N.~S.}\ \bibnamefont
  {Craigie}}\ and\ \bibinfo {author} {\bibfnamefont {H.}~\bibnamefont {Dorn}},\
  }\href {\doibase 10.1016/0550-3213(81)90372-2} {\bibfield  {journal}
  {\bibinfo  {journal} {Nucl. Phys. B}\ }\textbf {\bibinfo {volume} {185}},\
  \bibinfo {pages} {204} (\bibinfo {year} {1981})}\BibitemShut {NoStop}%
\bibitem [{\citenamefont {Dorn}(1986)}]{Dorn:1986dt}%
  \BibitemOpen
  \bibfield  {author} {\bibinfo {author} {\bibfnamefont {H.}~\bibnamefont
  {Dorn}},\ }\href {\doibase 10.1002/prop.19860340104} {\bibfield  {journal}
  {\bibinfo  {journal} {Fortsch. Phys.}\ }\textbf {\bibinfo {volume} {34}},\
  \bibinfo {pages} {11} (\bibinfo {year} {1986})}\BibitemShut {NoStop}%
\bibitem [{\citenamefont {Ji}\ \emph {et~al.}(2018)\citenamefont {Ji},
  \citenamefont {Zhang},\ and\ \citenamefont {Zhao}}]{Ji:2017oey}%
  \BibitemOpen
  \bibfield  {author} {\bibinfo {author} {\bibfnamefont {X.}~\bibnamefont
  {Ji}}, \bibinfo {author} {\bibfnamefont {J.-H.}\ \bibnamefont {Zhang}}, \
  and\ \bibinfo {author} {\bibfnamefont {Y.}~\bibnamefont {Zhao}},\ }\href
  {\doibase 10.1103/PhysRevLett.120.112001} {\bibfield  {journal} {\bibinfo
  {journal} {Phys. Rev. Lett.}\ }\textbf {\bibinfo {volume} {120}},\ \bibinfo
  {pages} {112001} (\bibinfo {year} {2018})},\ \Eprint
  {http://arxiv.org/abs/1706.08962} {arXiv:1706.08962 [hep-ph]} \BibitemShut
  {NoStop}%
\bibitem [{\citenamefont {Ishikawa}\ \emph {et~al.}(2017)\citenamefont
  {Ishikawa}, \citenamefont {Ma}, \citenamefont {Qiu},\ and\ \citenamefont
  {Yoshida}}]{Ishikawa:2017faj}%
  \BibitemOpen
  \bibfield  {author} {\bibinfo {author} {\bibfnamefont {T.}~\bibnamefont
  {Ishikawa}}, \bibinfo {author} {\bibfnamefont {Y.-Q.}\ \bibnamefont {Ma}},
  \bibinfo {author} {\bibfnamefont {J.-W.}\ \bibnamefont {Qiu}}, \ and\
  \bibinfo {author} {\bibfnamefont {S.}~\bibnamefont {Yoshida}},\ }\href
  {\doibase 10.1103/PhysRevD.96.094019} {\bibfield  {journal} {\bibinfo
  {journal} {Phys. Rev. D}\ }\textbf {\bibinfo {volume} {96}},\ \bibinfo
  {pages} {094019} (\bibinfo {year} {2017})},\ \Eprint
  {http://arxiv.org/abs/1707.03107} {arXiv:1707.03107 [hep-ph]} \BibitemShut
  {NoStop}%
\bibitem [{\citenamefont {Green}\ \emph {et~al.}(2018)\citenamefont {Green},
  \citenamefont {Jansen},\ and\ \citenamefont {Steffens}}]{Green:2017xeu}%
  \BibitemOpen
  \bibfield  {author} {\bibinfo {author} {\bibfnamefont {J.}~\bibnamefont
  {Green}}, \bibinfo {author} {\bibfnamefont {K.}~\bibnamefont {Jansen}}, \
  and\ \bibinfo {author} {\bibfnamefont {F.}~\bibnamefont {Steffens}},\ }\href
  {\doibase 10.1103/PhysRevLett.121.022004} {\bibfield  {journal} {\bibinfo
  {journal} {Phys. Rev. Lett.}\ }\textbf {\bibinfo {volume} {121}},\ \bibinfo
  {pages} {022004} (\bibinfo {year} {2018})},\ \Eprint
  {http://arxiv.org/abs/1707.07152} {arXiv:1707.07152 [hep-lat]} \BibitemShut
  {NoStop}%
\bibitem [{\citenamefont {Gao}\ \emph {et~al.}(2024)\citenamefont {Gao},
  \citenamefont {Liu},\ and\ \citenamefont {Zhao}}]{Gao:2023lny}%
  \BibitemOpen
  \bibfield  {author} {\bibinfo {author} {\bibfnamefont {X.}~\bibnamefont
  {Gao}}, \bibinfo {author} {\bibfnamefont {W.-Y.}\ \bibnamefont {Liu}}, \ and\
  \bibinfo {author} {\bibfnamefont {Y.}~\bibnamefont {Zhao}},\ }\href {\doibase
  10.1103/PhysRevD.109.094506} {\bibfield  {journal} {\bibinfo  {journal}
  {Phys. Rev. D}\ }\textbf {\bibinfo {volume} {109}},\ \bibinfo {pages}
  {094506} (\bibinfo {year} {2024})},\ \Eprint
  {http://arxiv.org/abs/2306.14960} {arXiv:2306.14960 [hep-ph]} \BibitemShut
  {NoStop}%
\bibitem [{\citenamefont {Zhao}(2024)}]{Zhao:2023ptv}%
  \BibitemOpen
  \bibfield  {author} {\bibinfo {author} {\bibfnamefont {Y.}~\bibnamefont
  {Zhao}},\ }\href {\doibase 10.1103/PhysRevLett.133.241904} {\bibfield
  {journal} {\bibinfo  {journal} {Phys. Rev. Lett.}\ }\textbf {\bibinfo
  {volume} {133}},\ \bibinfo {pages} {241904} (\bibinfo {year} {2024})},\
  \Eprint {http://arxiv.org/abs/2311.01391} {arXiv:2311.01391 [hep-ph]}
  \BibitemShut {NoStop}%
\bibitem [{\citenamefont {Hatta}\ \emph {et~al.}(2014)\citenamefont {Hatta},
  \citenamefont {Ji},\ and\ \citenamefont {Zhao}}]{Hatta:2013gta}%
  \BibitemOpen
  \bibfield  {author} {\bibinfo {author} {\bibfnamefont {Y.}~\bibnamefont
  {Hatta}}, \bibinfo {author} {\bibfnamefont {X.}~\bibnamefont {Ji}}, \ and\
  \bibinfo {author} {\bibfnamefont {Y.}~\bibnamefont {Zhao}},\ }\href {\doibase
  10.1103/PhysRevD.89.085030} {\bibfield  {journal} {\bibinfo  {journal} {Phys.
  Rev. D}\ }\textbf {\bibinfo {volume} {89}},\ \bibinfo {pages} {085030}
  (\bibinfo {year} {2014})},\ \Eprint {http://arxiv.org/abs/1310.4263}
  {arXiv:1310.4263 [hep-ph]} \BibitemShut {NoStop}%
\bibitem [{\citenamefont {Zhang}\ \emph {et~al.}(2024)\citenamefont {Zhang},
  \citenamefont {Huo}, \citenamefont {Ji}, \citenamefont {Schaefer},
  \citenamefont {Shi}, \citenamefont {Sun}, \citenamefont {Wang}, \citenamefont
  {Yang},\ and\ \citenamefont {Zhang}}]{Zhang:2024omt}%
  \BibitemOpen
  \bibfield  {author} {\bibinfo {author} {\bibfnamefont {K.}~\bibnamefont
  {Zhang}}, \bibinfo {author} {\bibfnamefont {Y.-K.}\ \bibnamefont {Huo}},
  \bibinfo {author} {\bibfnamefont {X.}~\bibnamefont {Ji}}, \bibinfo {author}
  {\bibfnamefont {A.}~\bibnamefont {Schaefer}}, \bibinfo {author}
  {\bibfnamefont {C.-J.}\ \bibnamefont {Shi}}, \bibinfo {author} {\bibfnamefont
  {P.}~\bibnamefont {Sun}}, \bibinfo {author} {\bibfnamefont {W.}~\bibnamefont
  {Wang}}, \bibinfo {author} {\bibfnamefont {Y.-B.}\ \bibnamefont {Yang}}, \
  and\ \bibinfo {author} {\bibfnamefont {J.-H.}\ \bibnamefont {Zhang}}
  (\bibinfo {collaboration} {Lattice Parton}),\ }\href {\doibase
  10.1103/PhysRevD.110.074505} {\bibfield  {journal} {\bibinfo  {journal}
  {Phys. Rev. D}\ }\textbf {\bibinfo {volume} {110}},\ \bibinfo {pages}
  {074505} (\bibinfo {year} {2024})},\ \Eprint
  {http://arxiv.org/abs/2405.14097} {arXiv:2405.14097 [hep-lat]} \BibitemShut
  {NoStop}%
\bibitem [{\citenamefont {Bollweg}\ \emph {et~al.}(2024)\citenamefont
  {Bollweg}, \citenamefont {Gao}, \citenamefont {Mukherjee},\ and\
  \citenamefont {Zhao}}]{Bollweg:2024zet}%
  \BibitemOpen
  \bibfield  {author} {\bibinfo {author} {\bibfnamefont {D.}~\bibnamefont
  {Bollweg}}, \bibinfo {author} {\bibfnamefont {X.}~\bibnamefont {Gao}},
  \bibinfo {author} {\bibfnamefont {S.}~\bibnamefont {Mukherjee}}, \ and\
  \bibinfo {author} {\bibfnamefont {Y.}~\bibnamefont {Zhao}},\ }\href {\doibase
  10.1016/j.physletb.2024.138617} {\bibfield  {journal} {\bibinfo  {journal}
  {Phys. Lett. B}\ }\textbf {\bibinfo {volume} {852}},\ \bibinfo {pages}
  {138617} (\bibinfo {year} {2024})},\ \Eprint
  {http://arxiv.org/abs/2403.00664} {arXiv:2403.00664 [hep-lat]} \BibitemShut
  {NoStop}%
\bibitem [{\citenamefont {Bollweg}\ \emph
  {et~al.}(2025{\natexlab{a}})\citenamefont {Bollweg}, \citenamefont {Gao},
  \citenamefont {He}, \citenamefont {Mukherjee},\ and\ \citenamefont
  {Zhao}}]{Bollweg:2025iol}%
  \BibitemOpen
  \bibfield  {author} {\bibinfo {author} {\bibfnamefont {D.}~\bibnamefont
  {Bollweg}}, \bibinfo {author} {\bibfnamefont {X.}~\bibnamefont {Gao}},
  \bibinfo {author} {\bibfnamefont {J.}~\bibnamefont {He}}, \bibinfo {author}
  {\bibfnamefont {S.}~\bibnamefont {Mukherjee}}, \ and\ \bibinfo {author}
  {\bibfnamefont {Y.}~\bibnamefont {Zhao}},\ }\href {\doibase
  10.1103/j3n6-8kxy} {\bibfield  {journal} {\bibinfo  {journal} {Phys. Rev. D}\
  }\textbf {\bibinfo {volume} {112}},\ \bibinfo {pages} {034501} (\bibinfo
  {year} {2025}{\natexlab{a}})},\ \Eprint {http://arxiv.org/abs/2504.04625}
  {arXiv:2504.04625 [hep-lat]} \BibitemShut {NoStop}%
\bibitem [{\citenamefont {Bollweg}\ \emph
  {et~al.}(2025{\natexlab{b}})\citenamefont {Bollweg}, \citenamefont {Gao},
  \citenamefont {Mukherjee},\ and\ \citenamefont {Zhao}}]{Bollweg:2025ecn}%
  \BibitemOpen
  \bibfield  {author} {\bibinfo {author} {\bibfnamefont {D.}~\bibnamefont
  {Bollweg}}, \bibinfo {author} {\bibfnamefont {X.}~\bibnamefont {Gao}},
  \bibinfo {author} {\bibfnamefont {S.}~\bibnamefont {Mukherjee}}, \ and\
  \bibinfo {author} {\bibfnamefont {Y.}~\bibnamefont {Zhao}},\ }\href {\doibase
  10.1103/tb2w-3tks} {\bibfield  {journal} {\bibinfo  {journal} {Phys. Rev.
  Lett.}\ }\textbf {\bibinfo {volume} {135}},\ \bibinfo {pages} {201901}
  (\bibinfo {year} {2025}{\natexlab{b}})},\ \Eprint
  {http://arxiv.org/abs/2505.18430} {arXiv:2505.18430 [hep-lat]} \BibitemShut
  {NoStop}%
\bibitem [{\citenamefont {Zwanziger}(1998)}]{Zwanziger:1998ez}%
  \BibitemOpen
  \bibfield  {author} {\bibinfo {author} {\bibfnamefont {D.}~\bibnamefont
  {Zwanziger}},\ }\href {\doibase 10.1016/S0550-3213(98)00031-5} {\bibfield
  {journal} {\bibinfo  {journal} {Nucl. Phys. B}\ }\textbf {\bibinfo {volume}
  {518}},\ \bibinfo {pages} {237} (\bibinfo {year} {1998})}\BibitemShut
  {NoStop}%
\bibitem [{\citenamefont {Baulieu}\ and\ \citenamefont
  {Zwanziger}(1999)}]{Baulieu:1998kx}%
  \BibitemOpen
  \bibfield  {author} {\bibinfo {author} {\bibfnamefont {L.}~\bibnamefont
  {Baulieu}}\ and\ \bibinfo {author} {\bibfnamefont {D.}~\bibnamefont
  {Zwanziger}},\ }\href {\doibase 10.1016/S0550-3213(99)00074-7} {\bibfield
  {journal} {\bibinfo  {journal} {Nucl. Phys. B}\ }\textbf {\bibinfo {volume}
  {548}},\ \bibinfo {pages} {527} (\bibinfo {year} {1999})},\ \Eprint
  {http://arxiv.org/abs/hep-th/9807024} {arXiv:hep-th/9807024} \BibitemShut
  {NoStop}%
\bibitem [{\citenamefont {Niegawa}(2006)}]{Niegawa:2006ey}%
  \BibitemOpen
  \bibfield  {author} {\bibinfo {author} {\bibfnamefont {A.}~\bibnamefont
  {Niegawa}},\ }\href {\doibase 10.1103/PhysRevD.74.045021} {\bibfield
  {journal} {\bibinfo  {journal} {Phys. Rev. D}\ }\textbf {\bibinfo {volume}
  {74}},\ \bibinfo {pages} {045021} (\bibinfo {year} {2006})},\ \Eprint
  {http://arxiv.org/abs/hep-th/0604142} {arXiv:hep-th/0604142} \BibitemShut
  {NoStop}%
\bibitem [{\citenamefont {Niegawa}\ \emph {et~al.}(2006)\citenamefont
  {Niegawa}, \citenamefont {Inui},\ and\ \citenamefont
  {Kohyama}}]{Niegawa:2006hg}%
  \BibitemOpen
  \bibfield  {author} {\bibinfo {author} {\bibfnamefont {A.}~\bibnamefont
  {Niegawa}}, \bibinfo {author} {\bibfnamefont {M.}~\bibnamefont {Inui}}, \
  and\ \bibinfo {author} {\bibfnamefont {H.}~\bibnamefont {Kohyama}},\ }\href
  {\doibase 10.1103/PhysRevD.74.105016} {\bibfield  {journal} {\bibinfo
  {journal} {Phys. Rev. D}\ }\textbf {\bibinfo {volume} {74}},\ \bibinfo
  {pages} {105016} (\bibinfo {year} {2006})},\ \Eprint
  {http://arxiv.org/abs/hep-th/0607207} {arXiv:hep-th/0607207} \BibitemShut
  {NoStop}%
\bibitem [{\citenamefont {Burgio}\ \emph {et~al.}(2012)\citenamefont {Burgio},
  \citenamefont {Schrock}, \citenamefont {Reinhardt},\ and\ \citenamefont
  {Quandt}}]{Burgio:2012ph}%
  \BibitemOpen
  \bibfield  {author} {\bibinfo {author} {\bibfnamefont {G.}~\bibnamefont
  {Burgio}}, \bibinfo {author} {\bibfnamefont {M.}~\bibnamefont {Schrock}},
  \bibinfo {author} {\bibfnamefont {H.}~\bibnamefont {Reinhardt}}, \ and\
  \bibinfo {author} {\bibfnamefont {M.}~\bibnamefont {Quandt}},\ }\href
  {\doibase 10.1103/PhysRevD.86.014506} {\bibfield  {journal} {\bibinfo
  {journal} {Phys. Rev. D}\ }\textbf {\bibinfo {volume} {86}},\ \bibinfo
  {pages} {014506} (\bibinfo {year} {2012})},\ \Eprint
  {http://arxiv.org/abs/1204.0716} {arXiv:1204.0716 [hep-lat]} \BibitemShut
  {NoStop}%
\bibitem [{\citenamefont {Belitsky}\ \emph {et~al.}(2004)\citenamefont
  {Belitsky}, \citenamefont {Ji},\ and\ \citenamefont
  {Yuan}}]{Belitsky:2003nz}%
  \BibitemOpen
  \bibfield  {author} {\bibinfo {author} {\bibfnamefont {A.~V.}\ \bibnamefont
  {Belitsky}}, \bibinfo {author} {\bibfnamefont {X.-d.}\ \bibnamefont {Ji}}, \
  and\ \bibinfo {author} {\bibfnamefont {F.}~\bibnamefont {Yuan}},\ }\href
  {\doibase 10.1103/PhysRevD.69.074014} {\bibfield  {journal} {\bibinfo
  {journal} {Phys. Rev. D}\ }\textbf {\bibinfo {volume} {69}},\ \bibinfo
  {pages} {074014} (\bibinfo {year} {2004})},\ \Eprint
  {http://arxiv.org/abs/hep-ph/0307383} {arXiv:hep-ph/0307383} \BibitemShut
  {NoStop}%
\bibitem [{\citenamefont {Collins}(2011)}]{Collins:2011zzd}%
  \BibitemOpen
  \bibfield  {author} {\bibinfo {author} {\bibfnamefont {J.}~\bibnamefont
  {Collins}},\ }\href {\doibase 10.1017/9781009401845} {\emph {\bibinfo {title}
  {{Foundations of Perturbative QCD}}}},\ Vol.~\bibinfo {volume} {32}\
  (\bibinfo  {publisher} {Cambridge University Press},\ \bibinfo {year}
  {2011})\BibitemShut {NoStop}%
\bibitem [{\citenamefont {Ji}\ \emph {et~al.}(2020)\citenamefont {Ji},
  \citenamefont {Liu},\ and\ \citenamefont {Liu}}]{Ji:2019ewn}%
  \BibitemOpen
  \bibfield  {author} {\bibinfo {author} {\bibfnamefont {X.}~\bibnamefont
  {Ji}}, \bibinfo {author} {\bibfnamefont {Y.}~\bibnamefont {Liu}}, \ and\
  \bibinfo {author} {\bibfnamefont {Y.-S.}\ \bibnamefont {Liu}},\ }\href
  {\doibase 10.1016/j.physletb.2020.135946} {\bibfield  {journal} {\bibinfo
  {journal} {Phys. Lett. B}\ }\textbf {\bibinfo {volume} {811}},\ \bibinfo
  {pages} {135946} (\bibinfo {year} {2020})},\ \Eprint
  {http://arxiv.org/abs/1911.03840} {arXiv:1911.03840 [hep-ph]} \BibitemShut
  {NoStop}%
\bibitem [{\citenamefont {Bazavov}\ \emph {et~al.}(2013)\citenamefont {Bazavov}
  \emph {et~al.}}]{MILC:2012znn}%
  \BibitemOpen
  \bibfield  {author} {\bibinfo {author} {\bibfnamefont {A.}~\bibnamefont
  {Bazavov}} \emph {et~al.} (\bibinfo {collaboration} {MILC}),\ }\href
  {\doibase 10.1103/PhysRevD.87.054505} {\bibfield  {journal} {\bibinfo
  {journal} {Phys. Rev. D}\ }\textbf {\bibinfo {volume} {87}},\ \bibinfo
  {pages} {054505} (\bibinfo {year} {2013})},\ \Eprint
  {http://arxiv.org/abs/1212.4768} {arXiv:1212.4768 [hep-lat]} \BibitemShut
  {NoStop}%
\bibitem [{\citenamefont {Symanzik}(1983)}]{Symanzik:1983dc}%
  \BibitemOpen
  \bibfield  {author} {\bibinfo {author} {\bibfnamefont {K.}~\bibnamefont
  {Symanzik}},\ }\href {\doibase 10.1016/0550-3213(83)90468-6} {\bibfield
  {journal} {\bibinfo  {journal} {Nucl. Phys. B}\ }\textbf {\bibinfo {volume}
  {226}},\ \bibinfo {pages} {187} (\bibinfo {year} {1983})}\BibitemShut
  {NoStop}%
\bibitem [{\citenamefont {Hasenfratz}\ and\ \citenamefont
  {Knechtli}(2001)}]{Hasenfratz:2001hp}%
  \BibitemOpen
  \bibfield  {author} {\bibinfo {author} {\bibfnamefont {A.}~\bibnamefont
  {Hasenfratz}}\ and\ \bibinfo {author} {\bibfnamefont {F.}~\bibnamefont
  {Knechtli}},\ }\href {\doibase 10.1103/PhysRevD.64.034504} {\bibfield
  {journal} {\bibinfo  {journal} {Phys. Rev. D}\ }\textbf {\bibinfo {volume}
  {64}},\ \bibinfo {pages} {034504} (\bibinfo {year} {2001})},\ \Eprint
  {http://arxiv.org/abs/hep-lat/0103029} {arXiv:hep-lat/0103029} \BibitemShut
  {NoStop}%
\bibitem [{\citenamefont {Sheikholeslami}\ and\ \citenamefont
  {Wohlert}(1985)}]{Sheikholeslami:1985ij}%
  \BibitemOpen
  \bibfield  {author} {\bibinfo {author} {\bibfnamefont {B.}~\bibnamefont
  {Sheikholeslami}}\ and\ \bibinfo {author} {\bibfnamefont {R.}~\bibnamefont
  {Wohlert}},\ }\href {\doibase 10.1016/0550-3213(85)90002-1} {\bibfield
  {journal} {\bibinfo  {journal} {Nucl. Phys. B}\ }\textbf {\bibinfo {volume}
  {259}},\ \bibinfo {pages} {572} (\bibinfo {year} {1985})}\BibitemShut
  {NoStop}%
\bibitem [{\citenamefont {Giusti}\ \emph {et~al.}(2001)\citenamefont {Giusti},
  \citenamefont {Paciello}, \citenamefont {Parrinello}, \citenamefont
  {Petrarca},\ and\ \citenamefont {Taglienti}}]{Giusti:2001xf}%
  \BibitemOpen
  \bibfield  {author} {\bibinfo {author} {\bibfnamefont {L.}~\bibnamefont
  {Giusti}}, \bibinfo {author} {\bibfnamefont {M.~L.}\ \bibnamefont
  {Paciello}}, \bibinfo {author} {\bibfnamefont {C.}~\bibnamefont
  {Parrinello}}, \bibinfo {author} {\bibfnamefont {S.}~\bibnamefont
  {Petrarca}}, \ and\ \bibinfo {author} {\bibfnamefont {B.}~\bibnamefont
  {Taglienti}},\ }\href {\doibase 10.1142/S0217751X01004281} {\bibfield
  {journal} {\bibinfo  {journal} {Int. J. Mod. Phys. A}\ }\textbf {\bibinfo
  {volume} {16}},\ \bibinfo {pages} {3487} (\bibinfo {year} {2001})},\ \Eprint
  {http://arxiv.org/abs/hep-lat/0104012} {arXiv:hep-lat/0104012} \BibitemShut
  {NoStop}%
\bibitem [{\citenamefont {Hudspith}(2015)}]{Hudspith:2014oja}%
  \BibitemOpen
  \bibfield  {author} {\bibinfo {author} {\bibfnamefont {R.~J.}\ \bibnamefont
  {Hudspith}} (\bibinfo {collaboration} {RBC, UKQCD}),\ }\href {\doibase
  10.1016/j.cpc.2014.10.017} {\bibfield  {journal} {\bibinfo  {journal}
  {Comput. Phys. Commun.}\ }\textbf {\bibinfo {volume} {187}},\ \bibinfo
  {pages} {115} (\bibinfo {year} {2015})},\ \Eprint
  {http://arxiv.org/abs/1405.5812} {arXiv:1405.5812 [hep-lat]} \BibitemShut
  {NoStop}%
\bibitem [{\citenamefont {Zhang}\ \emph {et~al.}(2025)\citenamefont {Zhang},
  \citenamefont {Grebe}, \citenamefont {Hackett}, \citenamefont {Wagman},\ and\
  \citenamefont {Zhao}}]{Zhang:2025hyo}%
  \BibitemOpen
  \bibfield  {author} {\bibinfo {author} {\bibfnamefont {R.}~\bibnamefont
  {Zhang}}, \bibinfo {author} {\bibfnamefont {A.~V.}\ \bibnamefont {Grebe}},
  \bibinfo {author} {\bibfnamefont {D.~C.}\ \bibnamefont {Hackett}}, \bibinfo
  {author} {\bibfnamefont {M.~L.}\ \bibnamefont {Wagman}}, \ and\ \bibinfo
  {author} {\bibfnamefont {Y.}~\bibnamefont {Zhao}},\ }\href {\doibase
  10.1103/6dh4-6k4t} {\bibfield  {journal} {\bibinfo  {journal} {Phys. Rev. D}\
  }\textbf {\bibinfo {volume} {112}},\ \bibinfo {pages} {L051502} (\bibinfo
  {year} {2025})},\ \Eprint {http://arxiv.org/abs/2501.00729} {arXiv:2501.00729
  [hep-lat]} \BibitemShut {NoStop}%
\bibitem [{\citenamefont {Reitinger}\ \emph {et~al.}(2026)\citenamefont
  {Reitinger}, \citenamefont {Sizmann}, \citenamefont {Sch{\"a}fer},
  \citenamefont {Zhang},\ and\ \citenamefont {Zhao}}]{Reitinger:2026hta}%
  \BibitemOpen
  \bibfield  {author} {\bibinfo {author} {\bibfnamefont {D.}~\bibnamefont
  {Reitinger}}, \bibinfo {author} {\bibfnamefont {T.}~\bibnamefont {Sizmann}},
  \bibinfo {author} {\bibfnamefont {A.}~\bibnamefont {Sch{\"a}fer}}, \bibinfo
  {author} {\bibfnamefont {R.}~\bibnamefont {Zhang}}, \ and\ \bibinfo {author}
  {\bibfnamefont {Y.}~\bibnamefont {Zhao}},\ }\href@noop {} {\  (\bibinfo
  {year} {2026})},\ \Eprint {http://arxiv.org/abs/2606.02447} {arXiv:2606.02447
  [hep-lat]} \BibitemShut {NoStop}%
\bibitem [{\citenamefont {Bali}\ \emph {et~al.}(2016)\citenamefont {Bali},
  \citenamefont {Lang}, \citenamefont {Musch},\ and\ \citenamefont
  {Sch{\"a}fer}}]{Bali:2016lva}%
  \BibitemOpen
  \bibfield  {author} {\bibinfo {author} {\bibfnamefont {G.~S.}\ \bibnamefont
  {Bali}}, \bibinfo {author} {\bibfnamefont {B.}~\bibnamefont {Lang}}, \bibinfo
  {author} {\bibfnamefont {B.~U.}\ \bibnamefont {Musch}}, \ and\ \bibinfo
  {author} {\bibfnamefont {A.}~\bibnamefont {Sch{\"a}fer}},\ }\href {\doibase
  10.1103/PhysRevD.93.094515} {\bibfield  {journal} {\bibinfo  {journal} {Phys.
  Rev. D}\ }\textbf {\bibinfo {volume} {93}},\ \bibinfo {pages} {094515}
  (\bibinfo {year} {2016})},\ \Eprint {http://arxiv.org/abs/1602.05525}
  {arXiv:1602.05525 [hep-lat]} \BibitemShut {NoStop}%
\bibitem [{\citenamefont {Gao}\ \emph {et~al.}(2026{\natexlab{a}})\citenamefont
  {Gao}, \citenamefont {Grebe}, \citenamefont {Hackett}, \citenamefont
  {Wagman}, \citenamefont {Zhang},\ and\ \citenamefont {Zhao}}]{keli2026}%
  \BibitemOpen
  \bibfield  {author} {\bibinfo {author} {\bibfnamefont {X.}~\bibnamefont
  {Gao}}, \bibinfo {author} {\bibfnamefont {A.~V.}\ \bibnamefont {Grebe}},
  \bibinfo {author} {\bibfnamefont {D.~C.}\ \bibnamefont {Hackett}}, \bibinfo
  {author} {\bibfnamefont {M.~L.}\ \bibnamefont {Wagman}}, \bibinfo {author}
  {\bibfnamefont {R.}~\bibnamefont {Zhang}}, \ and\ \bibinfo {author}
  {\bibfnamefont {Y.}~\bibnamefont {Zhao}},\ }\href@noop {} {\  (\bibinfo
  {year} {2026}{\natexlab{a}})},\ \bibinfo {note} {in preparation}\BibitemShut
  {NoStop}%
\bibitem [{\citenamefont {Wagman}(2025)}]{Wagman:2024rid}%
  \BibitemOpen
  \bibfield  {author} {\bibinfo {author} {\bibfnamefont {M.~L.}\ \bibnamefont
  {Wagman}},\ }\href {\doibase 10.1103/pcvc-734h} {\bibfield  {journal}
  {\bibinfo  {journal} {Phys. Rev. Lett.}\ }\textbf {\bibinfo {volume} {134}},\
  \bibinfo {pages} {241901} (\bibinfo {year} {2025})},\ \Eprint
  {http://arxiv.org/abs/2406.20009} {arXiv:2406.20009 [hep-lat]} \BibitemShut
  {NoStop}%
\bibitem [{\citenamefont {Hackett}\ and\ \citenamefont
  {Wagman}(2025)}]{Hackett:2024xnx}%
  \BibitemOpen
  \bibfield  {author} {\bibinfo {author} {\bibfnamefont {D.~C.}\ \bibnamefont
  {Hackett}}\ and\ \bibinfo {author} {\bibfnamefont {M.~L.}\ \bibnamefont
  {Wagman}},\ }\href {\doibase 10.1103/zjzt-rv86} {\bibfield  {journal}
  {\bibinfo  {journal} {Phys. Rev. D}\ }\textbf {\bibinfo {volume} {112}},\
  \bibinfo {pages} {054506} (\bibinfo {year} {2025})},\ \Eprint
  {http://arxiv.org/abs/2407.21777} {arXiv:2407.21777 [hep-lat]} \BibitemShut
  {NoStop}%
\bibitem [{\citenamefont {Gao}\ \emph {et~al.}(2026{\natexlab{b}})\citenamefont
  {Gao}, \citenamefont {He}, \citenamefont {Lin}, \citenamefont {Mukherjee},
  \citenamefont {Petreczky}, \citenamefont {Zhang},\ and\ \citenamefont
  {Zhao}}]{Gao:2026hix}%
  \BibitemOpen
  \bibfield  {author} {\bibinfo {author} {\bibfnamefont {X.}~\bibnamefont
  {Gao}}, \bibinfo {author} {\bibfnamefont {J.}~\bibnamefont {He}}, \bibinfo
  {author} {\bibfnamefont {J.}~\bibnamefont {Lin}}, \bibinfo {author}
  {\bibfnamefont {S.}~\bibnamefont {Mukherjee}}, \bibinfo {author}
  {\bibfnamefont {P.}~\bibnamefont {Petreczky}}, \bibinfo {author}
  {\bibfnamefont {R.}~\bibnamefont {Zhang}}, \ and\ \bibinfo {author}
  {\bibfnamefont {Y.}~\bibnamefont {Zhao}},\ }\href@noop {} {\  (\bibinfo
  {year} {2026}{\natexlab{b}})},\ \Eprint {http://arxiv.org/abs/2602.11283}
  {arXiv:2602.11283 [hep-lat]} \BibitemShut {NoStop}%
\bibitem [{\citenamefont {Su}\ \emph {et~al.}(2023)\citenamefont {Su},
  \citenamefont {Holligan}, \citenamefont {Ji}, \citenamefont {Yao},
  \citenamefont {Zhang},\ and\ \citenamefont {Zhang}}]{Su:2022fiu}%
  \BibitemOpen
  \bibfield  {author} {\bibinfo {author} {\bibfnamefont {Y.}~\bibnamefont
  {Su}}, \bibinfo {author} {\bibfnamefont {J.}~\bibnamefont {Holligan}},
  \bibinfo {author} {\bibfnamefont {X.}~\bibnamefont {Ji}}, \bibinfo {author}
  {\bibfnamefont {F.}~\bibnamefont {Yao}}, \bibinfo {author} {\bibfnamefont
  {J.-H.}\ \bibnamefont {Zhang}}, \ and\ \bibinfo {author} {\bibfnamefont
  {R.}~\bibnamefont {Zhang}},\ }\href {\doibase
  10.1016/j.nuclphysb.2023.116201} {\bibfield  {journal} {\bibinfo  {journal}
  {Nucl. Phys. B}\ }\textbf {\bibinfo {volume} {991}},\ \bibinfo {pages}
  {116201} (\bibinfo {year} {2023})},\ \Eprint
  {http://arxiv.org/abs/2209.01236} {arXiv:2209.01236 [hep-ph]} \BibitemShut
  {NoStop}%
\bibitem [{\citenamefont {Ji}\ \emph {et~al.}(2025)\citenamefont {Ji},
  \citenamefont {Liu}, \citenamefont {Su},\ and\ \citenamefont
  {Zhang}}]{Ji:2024hit}%
  \BibitemOpen
  \bibfield  {author} {\bibinfo {author} {\bibfnamefont {X.}~\bibnamefont
  {Ji}}, \bibinfo {author} {\bibfnamefont {Y.}~\bibnamefont {Liu}}, \bibinfo
  {author} {\bibfnamefont {Y.}~\bibnamefont {Su}}, \ and\ \bibinfo {author}
  {\bibfnamefont {R.}~\bibnamefont {Zhang}},\ }\href {\doibase
  10.1007/JHEP03(2025)045} {\bibfield  {journal} {\bibinfo  {journal} {JHEP}\
  }\textbf {\bibinfo {volume} {03}},\ \bibinfo {pages} {045} (\bibinfo {year}
  {2025})},\ \Eprint {http://arxiv.org/abs/2410.12910} {arXiv:2410.12910
  [hep-ph]} \BibitemShut {NoStop}%
\bibitem [{\citenamefont {Barry}\ \emph {et~al.}(2023)\citenamefont {Barry},
  \citenamefont {Gamberg}, \citenamefont {Melnitchouk}, \citenamefont {Moffat},
  \citenamefont {Pitonyak}, \citenamefont {Prokudin},\ and\ \citenamefont
  {Sato}}]{Barry:2023qqh}%
  \BibitemOpen
  \bibfield  {author} {\bibinfo {author} {\bibfnamefont {P.~C.}\ \bibnamefont
  {Barry}}, \bibinfo {author} {\bibfnamefont {L.}~\bibnamefont {Gamberg}},
  \bibinfo {author} {\bibfnamefont {W.}~\bibnamefont {Melnitchouk}}, \bibinfo
  {author} {\bibfnamefont {E.}~\bibnamefont {Moffat}}, \bibinfo {author}
  {\bibfnamefont {D.}~\bibnamefont {Pitonyak}}, \bibinfo {author}
  {\bibfnamefont {A.}~\bibnamefont {Prokudin}}, \ and\ \bibinfo {author}
  {\bibfnamefont {N.}~\bibnamefont {Sato}} (\bibinfo {collaboration} {Jefferson
  Lab Angular Momentum (JAM)}),\ }\href {\doibase 10.1103/PhysRevD.108.L091504}
  {\bibfield  {journal} {\bibinfo  {journal} {Phys. Rev. D}\ }\textbf {\bibinfo
  {volume} {108}},\ \bibinfo {pages} {L091504} (\bibinfo {year} {2023})},\
  \Eprint {http://arxiv.org/abs/2302.01192} {arXiv:2302.01192 [hep-ph]}
  \BibitemShut {NoStop}%
\bibitem [{\citenamefont {Vladimirov}(2019)}]{Vladimirov:2019bfa}%
  \BibitemOpen
  \bibfield  {author} {\bibinfo {author} {\bibfnamefont {A.}~\bibnamefont
  {Vladimirov}},\ }\href {\doibase 10.1007/JHEP10(2019)090} {\bibfield
  {journal} {\bibinfo  {journal} {JHEP}\ }\textbf {\bibinfo {volume} {10}},\
  \bibinfo {pages} {090} (\bibinfo {year} {2019})},\ \Eprint
  {http://arxiv.org/abs/1907.10356} {arXiv:1907.10356 [hep-ph]} \BibitemShut
  {NoStop}%
\bibitem [{\citenamefont {Cerutti}\ \emph {et~al.}(2023)\citenamefont
  {Cerutti}, \citenamefont {Rossi}, \citenamefont {Venturini}, \citenamefont
  {Bacchetta}, \citenamefont {Bertone}, \citenamefont {Bissolotti},\ and\
  \citenamefont {Radici}}]{Cerutti:2022lmb}%
  \BibitemOpen
  \bibfield  {author} {\bibinfo {author} {\bibfnamefont {M.}~\bibnamefont
  {Cerutti}}, \bibinfo {author} {\bibfnamefont {L.}~\bibnamefont {Rossi}},
  \bibinfo {author} {\bibfnamefont {S.}~\bibnamefont {Venturini}}, \bibinfo
  {author} {\bibfnamefont {A.}~\bibnamefont {Bacchetta}}, \bibinfo {author}
  {\bibfnamefont {V.}~\bibnamefont {Bertone}}, \bibinfo {author} {\bibfnamefont
  {C.}~\bibnamefont {Bissolotti}}, \ and\ \bibinfo {author} {\bibfnamefont
  {M.}~\bibnamefont {Radici}} (\bibinfo {collaboration} {MAP (Multi-dimensional
  Analyses of Partonic distributions)}),\ }\href {\doibase
  10.1103/PhysRevD.107.014014} {\bibfield  {journal} {\bibinfo  {journal}
  {Phys. Rev. D}\ }\textbf {\bibinfo {volume} {107}},\ \bibinfo {pages}
  {014014} (\bibinfo {year} {2023})},\ \Eprint
  {http://arxiv.org/abs/2210.01733} {arXiv:2210.01733 [hep-ph]} \BibitemShut
  {NoStop}%
\bibitem [{\citenamefont {Jiang}\ \emph {et~al.}(2024)\citenamefont {Jiang},
  \citenamefont {Shi}, \citenamefont {Chen}, \citenamefont {Gong},\ and\
  \citenamefont {Yang}}]{Jiang:2024lto}%
  \BibitemOpen
  \bibfield  {author} {\bibinfo {author} {\bibfnamefont {X.}~\bibnamefont
  {Jiang}}, \bibinfo {author} {\bibfnamefont {C.}~\bibnamefont {Shi}}, \bibinfo
  {author} {\bibfnamefont {Y.}~\bibnamefont {Chen}}, \bibinfo {author}
  {\bibfnamefont {M.}~\bibnamefont {Gong}}, \ and\ \bibinfo {author}
  {\bibfnamefont {Y.-B.}\ \bibnamefont {Yang}},\ }\href@noop {} {\  (\bibinfo
  {year} {2024})},\ \Eprint {http://arxiv.org/abs/2411.08461} {arXiv:2411.08461
  [hep-lat]} \BibitemShut {NoStop}%
\bibitem [{\citenamefont {Clark}\ \emph {et~al.}(2010)\citenamefont {Clark},
  \citenamefont {Babich}, \citenamefont {Barros}, \citenamefont {Brower},\ and\
  \citenamefont {Rebbi}}]{Clark:2009wm}%
  \BibitemOpen
  \bibfield  {author} {\bibinfo {author} {\bibfnamefont {M.~A.}\ \bibnamefont
  {Clark}}, \bibinfo {author} {\bibfnamefont {R.}~\bibnamefont {Babich}},
  \bibinfo {author} {\bibfnamefont {K.}~\bibnamefont {Barros}}, \bibinfo
  {author} {\bibfnamefont {R.~C.}\ \bibnamefont {Brower}}, \ and\ \bibinfo
  {author} {\bibfnamefont {C.}~\bibnamefont {Rebbi}} (\bibinfo {collaboration}
  {QUDA}),\ }\href {\doibase 10.1016/j.cpc.2010.05.002} {\bibfield  {journal}
  {\bibinfo  {journal} {Comput. Phys. Commun.}\ }\textbf {\bibinfo {volume}
  {181}},\ \bibinfo {pages} {1517} (\bibinfo {year} {2010})},\ \Eprint
  {http://arxiv.org/abs/0911.3191} {arXiv:0911.3191 [hep-lat]} \BibitemShut
  {NoStop}%
\bibitem [{\citenamefont {Avkhadiev}\ \emph {et~al.}(2024)\citenamefont
  {Avkhadiev}, \citenamefont {Shanahan}, \citenamefont {Wagman},\ and\
  \citenamefont {Zhao}}]{Avkhadiev:2024mgd}%
  \BibitemOpen
  \bibfield  {author} {\bibinfo {author} {\bibfnamefont {A.}~\bibnamefont
  {Avkhadiev}}, \bibinfo {author} {\bibfnamefont {P.~E.}\ \bibnamefont
  {Shanahan}}, \bibinfo {author} {\bibfnamefont {M.~L.}\ \bibnamefont
  {Wagman}}, \ and\ \bibinfo {author} {\bibfnamefont {Y.}~\bibnamefont
  {Zhao}},\ }\href {\doibase 10.1103/PhysRevLett.132.231901} {\bibfield
  {journal} {\bibinfo  {journal} {Phys. Rev. Lett.}\ }\textbf {\bibinfo
  {volume} {132}},\ \bibinfo {pages} {231901} (\bibinfo {year} {2024})},\
  \Eprint {http://arxiv.org/abs/2402.06725} {arXiv:2402.06725 [hep-lat]}
  \BibitemShut {NoStop}%
\end{thebibliography}
%

\end{document}